\documentclass[pdflatex,sn-mathphys-ay]{sn-jnl}

\usepackage{graphicx}%
\usepackage{multirow}%
\usepackage{amsmath,amssymb,amsfonts}%
\usepackage{amsthm}%
\usepackage{mathrsfs}%
\usepackage[title]{appendix}%
\usepackage{xcolor}%
\usepackage{textcomp}%
\usepackage{manyfoot}%
\usepackage{booktabs}%
\usepackage{algorithm}%
\usepackage{algorithmicx}%
\usepackage{algpseudocode}%
\usepackage{listings}%

\usepackage{subcaption}
\usepackage{mathtools}
\usepackage{hyperref}

\newcommand{\bew}{\begin{widetext}}
	\newcommand{\eew}{\end{widetext}}
\newcommand{\beq}{\begin{equation}}
	\newcommand{\eeq}{\end{equation}}
\newcommand{\bes}{\begin{subequations}}
	\newcommand{\ees}{\end{subequations}}
\newcommand{\mbs}[1]{\boldsymbol{#1}}
\newcommand{\mbf}[1]{\mathbf{#1}}
\newcommand{\msf}[1]{\mathsf{#1}}
\newcommand{\mrm}[1]{\mathrm{#1}}
\newcommand{\mcal}[1]{\mathcal{#1}}
\newcommand{\del}{\nabla}

\newcommand{\dsub}[1]{\partial_{#1}}

\newcommand{\abs}[1]{{\vert {#1} \vert}}
\newcommand{\norm}[1]{{\lVert {#1} \rVert}}

\newcommand{\bmat}[1]{\begin{bmatrix} #1 \end{bmatrix}}

\newcommand{\revtext}[1]{{\color{black} {#1}}}

\theoremstyle{thmstyleone}%
\theoremstyle{thmstyletwo}%

\theoremstyle{thmstylethree}%

\begin{document}

\title[Excited States Of Positronium]{Excited States Of Positronium From The Two Body Dirac Equations Of Constraint}


\author*[1]{\fnm{Robert} \sur{Johnson}}\email{robjohnson@alphawaveresearch.com}

%

\affil*[1]{\orgname{Alphawave Research}, \orgaddress{\street{29 Stanebrook Ct.}, \city{Jonesboro}, \postcode{30238}, \state{GA}, \country{USA}}}

%


\abstract{
The binding energies of the excited states of positronium are calculated using the two body Dirac equations of constraint formalism.  The results from nonperturbative evaluation are compared to those from perturbative evaluation.  Explicit forms of the coupled equations are presented in addition to the decoupled equations.  The equations for decoupled sates with J=0 are compared to an alternate formulation of the model.  Some misprints in the literature are identified and corrected.
}




\maketitle


\section{\label{sec:introduction}Introduction}
Recent investigations of the positronium spectrum~\cite{adkins2022precisionsprectroscopy1} have relied primarily on non-relativistic quantum electrodynamics (nrQED)~\cite{caswell1986effectivelagrangian,pineda1998potentialnrqed,pas2015anintrotorqed} for the interpretation of experimental observations.  However, classical electrodynamics is covariant by construction, as should be any quantum formulation.  An alternative theory has been presented in the literature, begun by Dirac~\cite{Dirac_1950,dirac1964lectures}, developed by Todorov~\cite{todorov1971PhysRevD.3.2351}, and pursued by various groups~\cite{vanalstine1982scalarinteractions,crater1983twobody} and \cite{sazdjian1985quantummechanicaltransform,sazdjian1986PhysRevD.33.3401,scott1992PhysRevA.45.4393}  This paper presents an implementation of that theory~\cite{crater1992PhysRevD.46.5117,crater2012PhysRevD.85.116005} for equal mass QED systems which has been released as open source code for the GNU Octave environment.

Many readers will be unfamiliar with this approach to bound states despite its relation to quantum field theory through the Bethe-Salpeter equation being long established~\cite{gellmann_low1951PhysRev.84.350}.  Its structure can be obtained from summation of one particle exchange diagrams~\cite{sazdjian1985quantummechanicaltransform,sazdjian1986PhysRevD.33.3401,crateretal1991covariantextrapolation,jallouli1997376}.  Recoil effects are described by a relativistic force balance constraint~\cite{crater1992PhysRevD.46.5117}, and the effective potential remains non-singular~\cite{crateretal1996singularityfree}.  The only effect missing is that of particle-antiparticle annihilation, \revtext{which affects only the states with zero orbital angular momentum}.  The interested reader is encouraged to follow its development in the literature for a historical perspective on its relation to other approaches to bound states in quantum field theory.

This paper is organized as follows.  In Section~\ref{sec:simple} we assess the numerical method for nonperturbative evaluation of bound state binding energies by replicating the results for a simple model~\cite{scott1992PhysRevA.45.4393}  In Section~\ref{sec:twobody} we review the two body equations of constraint presented by~\cite{crater2012PhysRevD.85.116005}  In Section~\ref{sec:eigenvalues} We demonstrate how we treat the RHS of the eigenvalue problem.  In Section~\ref{sec:coordinates} we show how various coordinate transformations are used to accomplish the numerical analysis.  In Section~\ref{sec:improvement} We show how an approximation made on the LHS of the eigenvalue problem is improved through iteration.  In Section~\ref{sec:perturbative} we replicate the perturbative results of~\cite{crater1992PhysRevD.46.5117} by identifying a misprint in the literature for the analytic formula.  In Section~\ref{sec:nonperturbative} we present the numerical results of the eigenvalue analysis and compare them to the perturbative values.  In Section~\ref{sec:comparison} we compare these results to those obtained from an improved Breit equation embodying the same physics.  In Section~\ref{sec:conclusion} we give our concluding remarks.

\section{\label{sec:simple}A Simple Example}
Let us start by assessing the numerical method for eigenvalue analysis applied to a simple model of the positronium system.  Naturally, we choose to work in natural units $\hbar \equiv 1$, $c \equiv 1$, and $m_e \equiv 1$, with electromagnetic coupling $\alpha$.  In these units, the first Bohr radius has value $r_B = 1 / \alpha$.  Our first model is for a single fermion in a Coulomb potential, and we can write Eqn.~(4.1) from~\cite{scott1992PhysRevA.45.4393} as \beq \label{eqn:ferm1simplJ}
\msf{H} \mbf{V} = \bmat{V(r) + E_0 & -\dsub{r} + (\kappa - 1) / r \\ \dsub{r} + (\kappa + 1) / r & V(r) - E_0} \bmat{g \\ f} = E \bmat{g \\ f} \;,
\eeq where $V(r) = -\alpha / r$ is the central potential, $E_0 = m_e c^2$ is the rest mass of the fermion, $\dsub{r} \equiv d/dr$ is the kinetic term, and $E$ is the total energy of the state.  The components of the eigenvector are $[g(r), f(r)] \equiv \mbf{V}^T$ such that $\norm{g} \gg \norm{f}$, and $\msf{H}$ is the Hamiltonian matrix.  The spin parameter $\kappa = L$ for $J = L - S$, and $\kappa = -(L+1)$ for $J = L + S$ for $S = 1/2$.  The kinetic term is approximated using finite differences, \beq
\dfrac{dv}{dr} \approx \dfrac{[1, -8, 0, 8, -1]}{12 \Delta_r} \bmat{v(k-2) \\ v(k-1) \\ v(k) \\ v(k+1) \\ v(k+2)}
\eeq when applied to vector $v$ discretely sampled over $r$ with spacing $\Delta_r$ such that $r_k \equiv k \Delta_r$.  Note that the boundary value $v(0) = 0$ is not stored explicitly but is considered when evaluating finite differences.  In other words, the finite difference stencil (selection of points) is allowed to slide one position further than the stored array, since the function value at that location is 0.  For a stencil of order 1, i.e. points $[-1, 0, 1]$, that is sufficient to account for the edge correction, but for a stencil of greater length one must modify the coefficients when approaching a finite boundary, i.e. $r \ge 0$ beyond which points do not exist.  The other boundary at $r = r_\mrm{max}$ does not require edge correction as function values beyond that location are treated as 0.  A convenient formula for that calculation is given by~\cite{taylor2016finitedifference} which we implement with some additional iterative improvement and symmetrization to improve numerical accuracy; details can be found in Appendix~\ref{sec:findiffcoef}.  In Table~\ref{tab:findif} we show the floating point coefficients for the edge corrected approximation above with unit step size.  All calculations are done in double precision.

\begin{table}
	\centering
	\begin{tabular}{|c|lllll|}
		\hline
		$k$ & \multicolumn{1}{c}{1} & \multicolumn{1}{c}{2} & \multicolumn{1}{c}{3} & \multicolumn{1}{c}{4} & \multicolumn{1}{c|}{5} \\
		\hline
		1 & -0.833333333333 & 1.500000000000 & -0.500000000000 & 0.083333333333 & 0.000000000000 \\ 
		2 & -0.666666666667 & 0.000000000000 & 0.666666666667 & -0.083333333333 & 0.000000000000 \\ 
		3 & 0.083333333333 & -0.666666666667 & 0.000000000000 & 0.666666666667 & -0.083333333333 \\ 
		\hline
	\end{tabular}
	\caption{First order derivative finite difference coefficients for a stencil of length 5 and unit step size $\Delta_r = 1$ at the left boundary $r = 0$}
	\label{tab:findif} 
\end{table}

Since there is suspicion of singularity of the wave function at the origin $r = 0$, we also consider a numerical evaluation over the coordinate $\rho = \log r$ using the natural base.  To effect the change of variable one simply has to modify the kinetic term to apply the finite difference coefficients to sampling over $\rho$ then use the transformation \beq
\dfrac{d}{dr} = \dfrac{1}{r} \dfrac{d}{d\rho} \;.
\eeq  Note that coordinate $\rho$ does not require the edge correction to the finite difference coefficients since the boundary values at $r = 0$ and $r = \infty$ are impossible to reach using a finite grid.  The expression for the potential is unchanged when expressed in terms of $r = \exp \rho$.  The Hamiltonian $\msf{H}$ is stored in sparse format, and to solve the eigensystem we rely on the ARPACK library.

\begin{table}
	\centering
	\begin{tabular}{|l|l||l|l|}
		\hline
		$\kappa$ & analytic & $\kappa$ & analytic \\
		$n$ & coordinate $r$ & $n$ & coordinate $r$ \\
		& coordinate $\rho$ &  & coordinate $\rho$ \\
		\hline
		-1 & 0.999973359973355264 & -1 & 0.999993339971160841 \\ 
		1 & 0.9999733599733560 & 2 & 0.9999933399711628 \\ 
		& 0.9999733599744481 & & 0.9999933399708602 \\ 
		\hline
		-1 & 0.999997039997039525 & -1 & 0.999998335002493105 \\ 
		3 & 0.9999970399970401 & 4 & 0.9999983350024934 \\ 
		& 0.9999970399969907 & & 0.9999983350024305 \\ 
		\hline
		-1 & 0.999998934403476272 & -2 & 0.999993340059872782 \\ 
		5 & 0.9999989344035400 & 2 & 0.9999933400598728 \\ 
		& 0.9999989344034567 & & 0.9999933400598728 \\ 
		\hline
		-2 & 0.999997040023324568 & -2 & 0.99999833501358211 \\ 
		3 & 0.9999970400233246 & 4 & 0.9999983350135822 \\ 
		& 0.9999970400233246 & & 0.9999983350135822 \\ 
		\hline
		-2 & 0.999998934409153828 & -3 & 0.999997040032086032 \\ 
		5 & 0.9999989344091323 & 3 & 0.9999970400320861 \\ 
		& 0.9999989344091538 & & 0.9999970400320861 \\ 
		\hline
		-3 & 0.999998335017278361 & -3 & 0.999998934411046314 \\ 
		4 & 0.9999983350172784 & 5 & 0.9999989344110389 \\ 
		& 0.9999983350172784 & & 0.9999989344110464 \\ 
		\hline
	\end{tabular}
	\caption{Estimates of $E / E_0$ for the one fermion system in a Coulomb potential}
	\label{tab:A} 
\end{table}

\begin{figure}
	\begin{center}
		\begin{subfigure}[t]{0.48\textwidth}
			\centering
			\includegraphics[width=\textwidth]{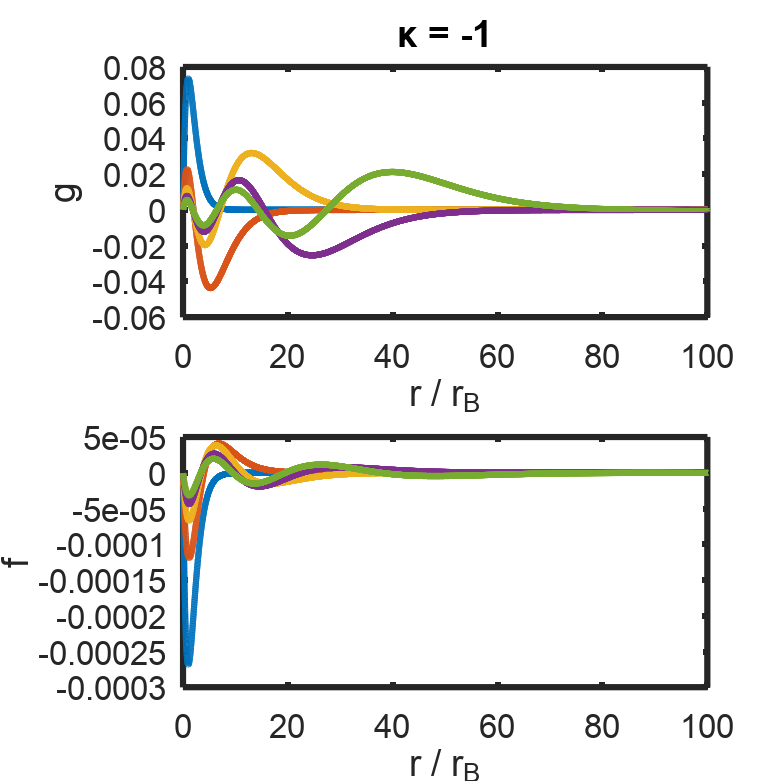}
			\caption{Eigenvectors over coordinate $r$}
			\label{fig:A1}
		\end{subfigure}
		\quad
		\begin{subfigure}[t]{0.48\textwidth}
			\centering
			\includegraphics[width=\textwidth]{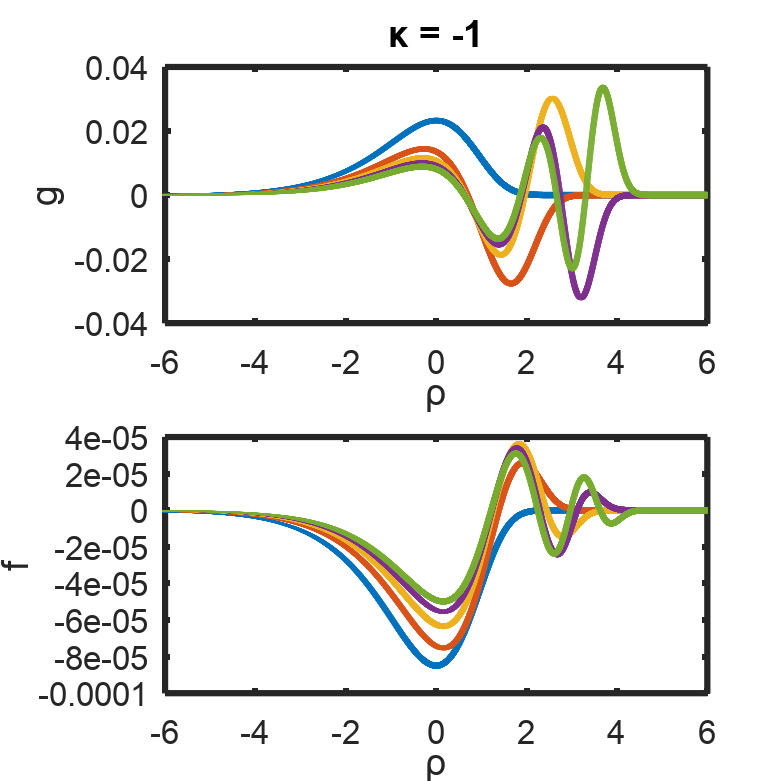}
			\caption{Eigenvectors over coordinate $\rho$}
			\label{fig:A4}
		\end{subfigure}
	\end{center}
	\caption{Eigenvectors for the one fermion system in a Coulomb potential}
	\label{fig:A}
\end{figure}

The analytic solution to Eqn.~(\ref{eqn:ferm1simplJ}) is well known~\cite{bethesalpeter1957oneandtwo}, \beq
E / E_0 = \left[ 1 + \left( \dfrac{\alpha Z}{n - |\kappa| + \sqrt{\kappa^2 - \alpha^2 Z^2}} \right)^2 \right]^{-1/2} \;,
\eeq where $Z = 1$ here and $n$ is the total quantum number, and can be calculated using arbitrary precision arithmetic.  In Table~\ref{tab:A} we compare values for the ratio $E / E_0$ given by the analytic formula to estimates over coordinates $r$ and $\rho$ using value $\alpha = 1 / 137$.  This table replicates Table~1 from~\cite{scott1992PhysRevA.45.4393}.  In Fig.~\ref{fig:A} we show some examples of the eigenvectors with $r$ in units of the Bohr radius $r_B$.  Note that the eigenvectors over $\rho$ are normalized to unity $\mbf{V}^T_n(\rho) \mbf{V}_n(\rho) = 1$ but are not orthogonal $\mbf{V}^T_n(\rho) \mbf{V}_{n'}(\rho) \neq 0$ since the matrix $\msf{H}(\rho)$ is not symmetric.

Next we recreate Table~2 from~\cite{scott1992PhysRevA.45.4393}.  We refer the reader to Eqns.~(2.8) through (2.12) therefrom for the model equations.  This model implements a two body Dirac equation with Coulomb potential and decouples into three sets of radial equations.  We perform the numerical estimates only over coordinate $\rho$.  This time we have no analytic formula for the energy, thus we take the value from the reference for comparison.  Results for the ratio $E / E_0$ where $E_0 = 2 m_e c^2$ are displayed in Table~\ref{tab:B}.

\begin{table}
	\centering
	\begin{tabular}{|llll|l|l|}
		\hline
		$n$ & $L$ & $S$ & $J$ & reference & coordinate $\rho$ \\
		\hline
		1 & 0 & 0 & 0 & 0.999993340148538880 & 0.9999933401485074 \\ 
		2 & 1 & 1 & 0 & 0.999998335009885854 & 0.9999983350098856 \\ 
		2 & 0 & 0 & 0 & 0.999998335024665402 & 0.9999983350246612 \\ 
		3 & 1 & 1 & 0 & 0.999999260005557351 & 0.9999992600055566 \\ 
		3 & 0 & 0 & 0 & 0.999999260009936472 & 0.9999992600099346 \\ 
		1 & 0 & 1 & 1 & 0.999993340148552498 & 0.9999933401485440 \\ 
		2 & 1 & 1 & 1 & 0.999998335013582156 & 0.9999983350135819 \\ 
		2 & 1 & 0 & 1 & 0.999998335017278391 & 0.9999983350172783 \\ 
		2 & 0 & 1 & 1 & 0.999998335024667154 & 0.9999983350246658 \\ 
		3 & 1 & 1 & 1 & 0.999999260006652550 & 0.9999992600066520 \\ 
		3 & 1 & 0 & 1 & 0.999999260007747730 & 0.9999992600077473 \\ 
		3 & 2 & 1 & 1 & 0.999999260007966753 & 0.9999992600079665 \\ 
		3 & 0 & 1 & 1 & 0.999999260009936994 & 0.9999992600099359 \\ 
		2 & 1 & 1 & 2 & 0.999998335020974567 & 0.9999983350209746 \\ 
		3 & 2 & 1 & 2 & 0.999999260008404825 & 0.9999992600084047 \\ 
		3 & 2 & 0 & 2 & 0.999999260008623860 & 0.9999992600086238 \\ 
		3 & 1 & 1 & 2 & 0.999999260008842892 & 0.9999992600088426 \\ 
		3 & 2 & 1 & 3 & 0.999999260009061929 & 0.9999992600090619 \\ 
		\hline
	\end{tabular}
	\caption{Estimates of $E / E_0$ for the two fermion system in a Coulomb potential}
	\label{tab:B} 
\end{table}

\clearpage

\section{\label{sec:twobody}Two Body Dirac Equations Of Constraint For Equal Mass QED}
Let us now turn our attention to the two body Dirac equations of constraint for equal mass QED systems.  We will be following closely the presentations found in \cite{crater1992PhysRevD.46.5117} and \cite{crater2012PhysRevD.85.116005}, and we encourage the reader to peruse those articles and references therein for a better understanding of how the model is derived.  We will be focusing on how the model is applied.

Our starting point will be the model equations given in Appendix~B of \cite{crater2012PhysRevD.85.116005}.  For a given value of $J$ we have a singlet state $u_0$ with $L = J,\; S = 0$ coupled to the member of the triplet $u_1$ with equal orbital angular momentum $L = J,\; S = 1$ through the equations \begin{align} 
\nonumber
	\left[ -\dfrac{d^2}{dr^2} + \dfrac{J(J+1)}{r^2} + 2 \epsilon_w A - A^2 + \Phi_D - 3 \Phi_{SS} \right] u_0 & \\
	 + 2 \sqrt{J(J+1)} ( \Phi_{SOD} - \Phi_{SOX} ) u_1 & = b_w^2 u_0 \;, \\
\nonumber
	\left[ -\dfrac{d^2}{dr^2} + \dfrac{J(J+1)}{r^2} + 2 \epsilon_w A - A^2 + \Phi_D - 2 \Phi_{SO} + \Phi_{SS} + 2 \Phi_T - 2 \Phi_{SOT} \right] u_1 & \\ + 2 \sqrt{J(J+1)} ( \Phi_{SOD} + \Phi_{SOX} ) u_0 & = b_w^2 u_1 \;.
\end{align}  The remaining states of the triplet, $u_+$ with $L = J - 1, S = 1$ and $u_-$ with $L = J + 1, S = 1$ (in the notation of the reference), are coupled by equations \begin{align} 
	\left[ -\dfrac{d^2}{dr^2} + \dfrac{J(J-1)}{r^2} + 2 \epsilon_w A - A^2 + \Phi_D + 2 (J-1) \Phi_{SO} + \Phi_{SS} + \dfrac{2(J-1)}{2J+1} ( \Phi_{SOT} - \Phi_T ) \right] u_+ & \\
	 + \dfrac{2\sqrt{J(J+1)}}{2J+1} \left[ 3 \Phi_T - 2 (J+2) \Phi_{SOT} \right] u_- = b_w^2 u_+ & \; \label{eqn:JPS1} \\ 
	\left[ -\dfrac{d^2}{dr^2} + \dfrac{(J+1)(J+2)}{r^2} + 2 \epsilon_w A - A^2 + \Phi_D - 2 (J+2) \Phi_{SO} + \Phi_{SS} + \dfrac{2(J+2)}{2J+1} ( \Phi_{SOT} - \Phi_T ) \right] u_- & \\
	 + \dfrac{2\sqrt{J(J+1)}}{2J+1} [ 3 \Phi_T + 2 (J-1) \Phi_{SOT} ] u_+ = b_w^2 u_- & \;. \label{eqn:JMS1}
\end{align}  We have already begun the process of simplification by omitting terms $2 m_w S + S^2$ in the equations above (where $S$ is a scalar potential associated with QCD) not required for the application to QED (to leading order).  To make sense of these equations, we must now write down expressions for all of the terms above, under the assumption that $m_2 = m_1$.

Let us begin with the kinematic variables~\cite{todorov1971PhysRevD.3.2351}.  For equal mass systems, we have the relations $m_w = m_1^2 / w$, $\epsilon_w = (w^2 - 2 m_1^2) / 2 w$, and $b_w^2 = \epsilon_w^2 - m_w^2$, where $m_w$ and $\epsilon_w$ are the reduced mass and effective energy of the two fermion system, and $b_w^2 = w^2/4-m_1^2$ is the squared relative momentum.  The relativistic energy of the state is given by $w$ and is the value for which we solve the equations.

Next let us tackle the expressions comprising the quasipotential $\Phi$.  These expressions come directly from Eqn.~(A28) in~\cite{crater2012PhysRevD.85.116005}, and the interpretation of the subscripts can be found in~\cite{crater1992PhysRevD.46.5117} \revtext{below its Eqn.~(4.15)}.  Explicitly, we have \begin{align}
	\Phi_T = & [ n(r) + (3 \mcal{F}' - \mcal{K}' + 3 / r) (\sinh 2 \mcal{K}) / r + (\mcal{F}' - 3 \mcal{K}' + 1 / r) (\cosh 2 \mcal{K} - 1) / r \nonumber \\
	& + 2 \mcal{F}' \mcal{K}' - \del^2 \mcal{K} ] / 3 \;, \\
	\Phi_{SS} = & [ 3 k(r) + 2 (\mcal{K}' \sinh 2 \mcal{K}) / r - 2 (\mcal{F}' + 1/r) (\cosh 2 \mcal{K} - 1) / r + 2 \mcal{F}' \mcal{K}' - \del^2 \mcal{K} ] / 3 \;, \\
	\Phi_D = & m(r) + 2 (\mcal{K}' \sinh 2 \mcal{K}) / r - 2 (\mcal{F}' + 1/r) (\cosh 2 \mcal{K} - 1) / r + \mcal{F}'^2 + \mcal{K}'^2 - \del^2 \mcal{F} \;, \\
	\Phi_{SO} = & -\mcal{F}' / r - (\mcal{F}' + 1/r) (\cosh 2 \mcal{K} - 1) / r + (\mcal{K}' \sinh 2 \mcal{K}) / r \;, \label{eqn:phiSO} \\
	\Phi_{SOT} = & (\mcal{F}' + 1/r) (\sinh 2 \mcal{K}) / r - \mcal{K}' (\cosh 2 \mcal{K} - 1) / r - \mcal{K}' / r \;, \\
	\Phi_{SOD} = & l' \cosh 2 \mcal{K} - q' \sinh 2 \mcal{K} \;, \\
	\Phi_{SOX} = & q' \cosh 2 \mcal{K} + l' \sinh 2 \mcal{K} \;,
\end{align} where $X' \equiv dX / dr$.  Note that there has recently been a change of sign in the expression for $\Phi_{SOX}$ \revtext{in~\cite{crater2012PhysRevD.85.116005} relative to~\cite{crater2009PhysRevD.79.034011,crater2010PhysRevD.82.094020}} and that we omit a factor of 2 from the expression for $\Phi_{SO}$ for reasons to be discussed later.

The terms $\Phi_{SOD}$ and $\Phi_{SOX}$ are irrelevant for the equal mass case because $l'$ and $q'$ are 0 when $m_2 = m_1$ according to Eqn.~(A29) from~\cite{crater2012PhysRevD.85.116005}.  The remainder of Eqn.~(A29) reads \begin{align}
	n(r) = & \del^2 \mcal{K} - \del^2 \mcal{G} / 2 + 3 (\mcal{G} - 2 \mcal{K})' / 2 r + \mcal{F}' \mcal{G}' - 2 \mcal{F}' \mcal{K}' \;, \\
	k(r) = & \del^2 (\mcal{G} + \mcal{K}) / 3 - \mcal{G}'^2 / 2 - 2 \mcal{F}' (\mcal{G} + \mcal{K})' / 3 \;, \\
	m(r) = & -\del^2 \mcal{G} / 2 + 3 \mcal{G}'^2 / 4 - \mcal{K}'^2 + \mcal{G}' \mcal{F}' \;,
\end{align} \revtext{where we note the expression for $n(r)$ has lost a factor of 1/3 relative to~\cite{crater2009PhysRevD.79.034011,crater2010PhysRevD.82.094020}.}  Because of that cancellation, the equations for $u_0$ and $u_1$ decouple (as well as for unequal masses when $J = 0$).  The equations for $u_+$ and $u_-$ also decouple when $J = 0$; for this special case, the equations for $u_1$ and $u_+$ describe unphysical states and may be discarded.

The next step is to implement the relations reduced to the case of vector (electromagnetic) interactions describing QED, explicitly \begin{align}
	\mcal{F}' = & - 3 \mcal{G}' / 2 \;, \hspace{1em} \mcal{K}' = \mcal{G}' / 2 \;, \hspace{1em} \mcal{G}' = A' / (w - 2 A) \;, \\
	\del^2 \mcal{F} = & - 3 \del^2 \mcal{G} / 2 \;, \hspace{1em} \del^2 \mcal{K} = \del^2 \mcal{G} / 2 \;, \hspace{1em} \del^2 \mcal{G} = (w - 2 A)^{-1} \del^2 A + 2 \mcal{G}'^2 \;, 
\end{align} and $A = - \alpha / r$ thus $A' = \alpha / r^2$.  The potential $A$ is related to the reduced charge density $\rho$ through the usual relation $\del^2 A(\mbf{r}) = 4 \pi \alpha \rho(\mbf{r})$, and the reduced charge density is treated as a point source at the origin $\rho(\mbf{r}) \rightarrow \delta(\mbf{r})$.  In~\cite{crater2012PhysRevD.85.116005} the effect of considering $\rho(\mbf{r})$ to be a distributed function is explored.  Our wave function has already absorbed a radial coordinate $u = r \psi$ which should be sufficient to negate the contribution of the delta function.

Finally, we need to have expressions for the hyperbolic cosine and sine terms, for which Eqn.~(B9) in~\cite{crater2012PhysRevD.85.116005} has an extra $-$ sign.  For the case of equal mass QED, the correct expressions are \begin{align}
	\cosh 2 \mcal{K} = \cosh \mcal{G} & = \dfrac{1}{2} \left[ (1 - 2 A / w)^{-1/2} + (1 - 2 A / w)^{1/2} \right] \;, \\
	\sinh 2 \mcal{K} = \sinh \mcal{G} & = \dfrac{1}{2} \left[ (1 - 2 A / w)^{-1/2} - (1 - 2 A / w)^{1/2} \right] \;, \label{eqn:sinh2K}
\end{align}  where $\mcal{G} = - \log (1 - 2 A / w)^{1/2}$.  More general expressions are found in~\cite{crater2012PhysRevD.85.116005}, \revtext{wherein Eqn.~(A27) $\mcal{K} = \mcal{G} / 2$ when $\mcal{L} = 0$.  The appearance of the $-$ sign in Eqn.~(B9) may have resulted when the authors consulted~\cite{crater2004PhysRevD.70.034026} wherein Eqn. (2.23) calls for $\sinh \mcal{J} = - \sinh \mcal{G}$.}

\subsection{Decoupled States $\prescript{1}{}J_J$, $\prescript{3}{}J_J$, and $\prescript{3}{}P_0$}
For the equal mass case $m_2 = m_1$, the states $\prescript{1}{}J_J$ and $\prescript{3}{}J_J$ decouple, as does the state $\prescript{3}{}P_0$ because it has $J = 0$.  Reduced expressions for those equations are given in~\cite{crater2012PhysRevD.85.116005}.  Let us now review these expressions.

Starting with the singlet case $L = J,\; S = 0$, the equation for $u_0$ reads \beq
\left[ -\dfrac{d^2}{dr^2} + \dfrac{J(J+1)}{r^2} + 2 \epsilon_w A - A^2 + \Phi_D - 3 \Phi_{SS} \right] u_0 = b_w^2 u_0 \;,
\eeq and the two terms of the quasipotential sum to 0, as \bes \begin{align} \label{eqn:phiJLS0}
	3 \Phi_{SS} & = (w - 2 A)^{-1} \del^2 A + 7 \mcal{G}'^2 / 2 + \mcal{G}' [ \sinh 2 \mcal{K} + 3 (\cosh 2 \mcal{K} - 1) ] / r - 2 (\cosh 2 \mcal{K} - 1) / r^2 \;, \\
	& = \Phi_D \;,
\end{align} \ees which for states $\prescript{1}{}J_J$ leaves us with the equation \beq \label{eqn:JLS0}
\left[ -\dfrac{d^2}{dr^2} + \dfrac{J(J+1)}{r^2} - \dfrac{2 \epsilon_w  \alpha}{r} - \dfrac{\alpha^2}{r^2} \right] u_0 = b_w^2 u_0 \;.
\eeq  

Similarly, the quasipotential for the case $L = J,\; S = 1$ contains no dependence on $J$, as \bes \begin{align} \label{eqn:phiJLS1}
	\Phi_D - 2 \Phi_{SO} + \Phi_{SS} + 2 \Phi_T - 2 \Phi_{SOT} & = \del^2 \mcal{G} + \mcal{G}'^2 - 2 \mcal{G}' / r \;, \\
	& = \dfrac{\del^2 A}{w - 2 A} + \dfrac{3 A'^2}{(w - 2 A)^2} - \dfrac{2 A'}{r (w - 2 A)} \;, \\
	& = \dfrac{4 \pi \alpha r \delta(\mbf{r})}{2 \alpha + r w} - \dfrac{\alpha (\alpha + 2 r w)}{r^2 (2 \alpha + r w)^2} \;,
\end{align} \ees thus for states $\prescript{3}{}J_J$ the equation for $u_1$ is \beq \label{eqn:JLS1}
\left[ -\dfrac{d^2}{dr^2} + \dfrac{J(J+1)}{r^2} - \dfrac{2 \epsilon_w  \alpha}{r} - \dfrac{\alpha^2}{r^2} - \dfrac{\alpha (\alpha + 2 r w)}{r^2 (2 \alpha + r w)^2} \right] u_1 = b_w^2 u_1
\eeq when the contribution from $r \delta(\mbf{r})$ is neglected.  Note when examining~\cite{crater2012PhysRevD.85.116005}  \revtext{that the $-$ signs in Eqn.~(B20) are inconsistent in their intermediate expression in terms of $\mcal{G}$} and that a factor of $1/2$ is introduced \revtext{to the quasipotential terms in Eqn.~(B22) which does not appear in the main text Eqn.~(24).}  We believe that Eqn.~(\ref{eqn:phiJLS1}) above is correct when the factor of 2 is not applied to the expression for $\Phi_{SO}$ in Eqn.~(\ref{eqn:phiSO}), \revtext{with further discussion in Appendix~\ref{sec:phisofactor2}}.

Finally, for the case $L = 1,\; S = 1,\; J = 0$ we have the quasipotential expression \bes \begin{align} \label{eqn:phi3P0}
	\Phi_D - 4 \Phi_{SO} + \Phi_{SS} - 4 \Phi_T + 4 \Phi_{SOT} & = \dfrac{2 \del^2 A}{w - 2 A} - \dfrac{8 A'}{r (w - 2 A)} + \dfrac{8 A'^2}{(w - 2 A)^2} \;, \\
	& = \dfrac{8 \pi \alpha r \delta(\mbf{r})}{2 \alpha + r w} - \dfrac{8 \alpha (\alpha + r w)}{r^2 (2 \alpha + r w)^2} \;,
\end{align} \ees which combined with the centrifugal term \beq
\dfrac{2}{r^2} - \dfrac{8 \alpha (\alpha + r w)}{r^2 (2 \alpha + r w)^2} = \dfrac{2 w^2}{(2 \alpha + r w)^2}
\eeq yields the corresponding equation for state $\prescript{3}{}P_0$, \beq \label{eqn:3P0}
\left[ -\dfrac{d^2}{dr^2} + \dfrac{2 w^2}{(2 \alpha + r w)^2} - \dfrac{2 \epsilon_w  \alpha}{r} - \dfrac{\alpha^2}{r^2} \right] u_- = b_w^2 u_- \;,
\eeq under the same condition $r \delta(\mbf{r}) \rightarrow 0$ \revtext{and $\Phi_{SO}$ as in Eqn.~(\ref{eqn:phiSO})}.  Note that Eqns.~(\ref{eqn:JLS0}), (\ref{eqn:JLS1}), and (\ref{eqn:3P0}) are in agreement with Eqns.~(23), (24), and (25) from the main text of~\cite{crater2012PhysRevD.85.116005} and are unaffected by the sign of Eqn.~(\ref{eqn:sinh2K}) above.

\subsection{Coupled States $\prescript{3}{}L_{L+1}$ and $\prescript{3}{}L_{L-1}$}
For the coupled states $\prescript{3}{}L_{L+1}$ and $\prescript{3}{}L_{L-1}$ there is not much simplification to be found.  One can work with the combined expressions \beq
\Phi_D + \Phi_{SS} = \dfrac{4 \del^2 A}{3 (w - 2 A)} + \dfrac{4 A' [ \sinh 2 \mcal{K} + 3 (\cosh 2 \mcal{K} - 1) ]}{3 r (w - 2 A)} + \dfrac{14 A'^2}{3 (w - 2 A)^2} - \dfrac{8 (\cosh 2 \mcal{K} - 1)}{3 r^2} \;, 
\eeq and \beq
\Phi_{SOT} - \Phi_T = \dfrac{\del^2 A}{6 (w - 2 A)} + \dfrac{A' [ \sinh 2 \mcal{K} + 3 (\cosh 2 \mcal{K} - 2) ]}{6 r (w - 2 A)} + \dfrac{5 A'^2}{6 (w - 2 A)^2} - \dfrac{\cosh 2 \mcal{K} - 1}{3 r^2} \;,
\eeq but because of the spin structure's dependence on $J$, one also needs expressions for $\Phi_{SO}$, $\Phi_{SOT}$, and $\Phi_T$ independently, \begin{align}
	\Phi_{SO} = & \dfrac{A' (\sinh 2 \mcal{K} + 3 \cosh 2 \mcal{K})}{2 r (w - 2 A)} - \dfrac{\cosh 2 \mcal{K} - 1}{r^2} \;, \\
	\Phi_{SOT} = & - \dfrac{A' (3 \sinh 2 \mcal{K} + \cosh 2 \mcal{K})}{2 r (w - 2 A)} + \dfrac{\sinh 2 \mcal{K}}{r^2} \;, \\
	\Phi_T = & - \dfrac{\del^2 A}{6 (w - 2 A)} - \dfrac{A' [ 5 \sinh 2 \mcal{K} + 3 (\cosh 2 \mcal{K} - 1) ]}{3 r (w - 2 A)} - \dfrac{5 A'^2}{6 (w - 2 A)^2} \nonumber \\
	& + \dfrac{3 \sinh 2 \mcal{K} +  \cosh 2 \mcal{K} - 1}{3 r^2} \;.
\end{align}  

These expressions are sensitive to the sign of Eqn.~(\ref{eqn:sinh2K}) above, \revtext{noting that Appendix~B of~\cite{crater2012PhysRevD.85.116005} does not give explicitly the reduction of the quasipotential in the coupled state equations for comparison}.  The coupled equations can be written in matrix format, \beq
\bmat{H_{++} & H_{+-} \\ H_{-+} & H_{--}} \bmat{u_+ \\ u_-} = b_w^2 \bmat{u_+ \\ u_-} \;,
\eeq where the expressions for the matrix components are found in Eqns.~(\ref{eqn:JPS1}) and (\ref{eqn:JMS1}).

\section{\label{sec:eigenvalues}Eigenvalue Analysis}
Looking now at the RHS of our eigenvalue problem, we find the term $b_w^2 u$, where $u$ could be either the decoupled or coupled wave function.  Expanding that expression gives \beq
b_w^2 u = (w^2 / 4 - m_1^2) u = (w^2 / 4) [1 - (2 m_1 / w)^2] u \equiv (w / 2)^2 \tilde{b}_w^2 u \;.
\eeq  The factor $(w/2)^2$ is moved over to the LHS where its reciprocal multiplies the kinetic and quasipotential terms, leaving $\tilde{b}_w^2 u$ on the RHS.  The term in $\tilde{b}_w^2$ with value 1 is recognized as an offset to the eigenvalue and may also be moved to the LHS.  That leaves on the RHS the expression $(\tilde{b}_w^2 - 1) \equiv d \simeq -1$ as the eigenvalue.  Multiple eigenvalues can be combined into a single diagonal matrix $\msf{D}$ as long as the Hamiltonian matrix is equivalent for each, $\msf{H} \msf{V} = \msf{V} \msf{D}$, which is the form of solution returned by the ARPACK library.  The algorithm employed excels at resolving the first eigenvector, but to improve the accuracy we iterate the solution method twice on each eigenvector independently.  Knowing a value of $d$, we can find the energy $w$ of the state using the formula $w = 2 m_1 / \sqrt{-d}$, since $d = - (2 m_1 / w)^2$.

\section{\label{sec:coordinates}Coordinate Transformations}
To put these equations into a form amenable to numerical analysis, one commonly employs a transformation to a coordinate proportional to the radius, e.g. $\tilde{r} = r \alpha \epsilon_w \propto r/r_B$.  While that transformation simplifies Eqn.~(\ref{eqn:JLS0}), it does not help the remaining equations on account of all the factors of $w$ appearing on the LHS.  Inspecting the expression \beq
\mcal{G} = - \log (1 - 2 A / w)^{1/2} = - \log (1 + 2 \alpha / r w)^{1/2} \;,
\eeq ones sees that the natural transformation is to a coordinate $y = r w / 2 \alpha \in [0, \infty]$.  While $r \alpha = r / r_B$ is a natural scaling for human understanding, that ratio turns the equations into polynomials in $\alpha$ subject to round-off error.  We will also make use of coordinates $z = \log y \in [-\infty, \infty]$ and $x = (1 + s / y)^{-1} \in [0, 1]$, where $s$ is a scaling factor that controls the mapping from $y$ to $x$.

Let us first address the kinetic term which all equations share.  Repeated application of the chain rule yields \beq
\dfrac{d^2}{dr^2} = \left( \dfrac{dy}{dr} \right)^2 \dfrac{d^2}{dy^2} + \left( \dfrac{dy}{dr} \right) \left( \dfrac{d}{dy} \dfrac{dy}{dr} \right) \dfrac{d}{dy} = \dfrac{w^2}{4 \alpha^2} \dfrac{d^2}{dy^2} \;.
\eeq  Similarly, we have the expressions \bes \begin{align}
	\dfrac{d^2}{dy^2} & = \dfrac{1}{y^2} \left( \dfrac{d^2}{dz^2} - \dfrac{d}{dz} \right) \;, \\
	& = \dfrac{s}{(y+s)^3} \left[ \dfrac{s}{y+s} \dfrac{d^2}{dx^2} - 2 \dfrac{d}{dx} \right] \;,
\end{align} \ees and using the substitution $x(r) = (1 + 2 \alpha s / r w)^{-1}$ such that $r(x) = 2 \alpha s x / w (1 - x)$, we have \beq
\dfrac{d^2}{dr^2} = \dfrac{w^2 (1 - x)^3}{4 \alpha^2 s^2} \left[ (1 - x) \dfrac{d^2}{dx^2} - 2 \dfrac{d}{dx} \right] \;.
\eeq  Note that coordinate $x$ requires edge correction to the finite difference approximation at both ends of the axis, while coordinate $y$ does at only one.

We can put these coordinate axes on an equal footing by introducing the scale factor $s$ to the $y$ and $z$ coordinates.  We do that by defining a standard coordinate axis $\tilde{y} \equiv \tilde{y}_n = n / N$ with $N$ stored locations (since $\tilde{y}_0$ is implicit), then applying scale factor $s$ such that $y = s \tilde{y}$.  Similarly, if $\tilde{z} \in [\tilde{z}_1, \tilde{z}_N]$ is a standard axis, then it can be shifted by an amount $z = \log s + \tilde{z}$.  For coordinate $x$, the standard axis with $N$ points is $\tilde{x} \equiv \tilde{x}_n = n / (N+1)$, and we work with a variable $y = s \tilde{x} / (1 - \tilde{x})$.  After accounting for the effect of the coordinate transformation on the kinetic term, the remaining terms in the potential can be evaluated in terms of $y = y(\tilde{x}) = y(\tilde{y}) = y(\tilde{z})$ given scale factor $s$.

\subsection{Decoupled Equations}
Let us start by expanding the factor $\epsilon_w$ in Eqn.~(\ref{eqn:JLS0}).  For state vector $u_0$ the potential $V_0$ may be written \bes \begin{align}
	V_0(r) & = \dfrac{J(J+1)}{r^2} - \dfrac{\alpha^2}{r^2} - \dfrac{2 \epsilon_w  \alpha}{r} \;, \\
	& = \dfrac{J(J+1)}{r^2} - \dfrac{\alpha^2}{r^2} - \dfrac{w \alpha}{r} + \dfrac{2 m_1^2 \alpha}{r w} \;,
\end{align} \ees and we apply the coefficient $4/w^2$ from the eigenvalue analysis to form $\tilde{V}_0 = (4/w^2) V_0$.  After the coordinate transformation, we have the expression \beq
\tilde{V}_0(y) = \dfrac{J (J+1)}{\alpha^2 y^2} - \dfrac{2 y + 1}{y^2} + \dfrac{4 m_1^2}{w^2 y}
\eeq and a final equation of \beq
\left[ - \alpha^{-2} \dfrac{d^2}{dy^2} + \tilde{V}_0(y) - 1 \right] u_0 = d u_0 \;.
\eeq  Similarly, we find for state vector $u_1$ the expression $\tilde{V}_1$ to be \beq
\tilde{V}_1(y) = \dfrac{J (J+1) (y+1)^2 - y - 1/4}{\alpha^2 y^2 (y+1)^2} - \dfrac{2 y + 1}{y^2} + \dfrac{4 m_1^2}{w^2 y} \;,
\eeq and for the special case of $\prescript{3}{}P_0$ we have \beq
\tilde{V}_-(y) = \dfrac{2}{\alpha^2 (y+1)^2} - \dfrac{2 y + 1}{y^2} + \dfrac{4 m_1^2}{w^2 y} \;.
\eeq

\subsection{Coupled Equations}
For the coupled equations, let us isolate the expressions according to \bes \begin{align}
	\tilde{\msf{H}} & = \tilde{\msf{K}} + \tilde{\msf{V}} - \msf{I} \;, \\
	& = \bmat{\tilde{K}_{++} & 0 \\ 0 & \tilde{K}_{--}} + \bmat{\tilde{V}_{++} & \tilde{V}_{+-} \\ \tilde{V}_{-+} & \tilde{V}_{--}} - \bmat{1 & 0 \\ 0 & 1} \;,
\end{align} \ees where $\tilde{\msf{K}}$ are the kinetic terms, $\tilde{\msf{V}}$ are the potential terms, and where the identity matrix $\msf{I}$ comes from the eigenvalue analysis.  The kinetic terms are the same as above, $\tilde{K}_{++} = \tilde{K}_{--} = - \alpha^{-2} d^2 / dy^2$, and the potential terms share a common denominator, which we factor as \beq
\bmat{\tilde{V}_{++} & \tilde{V}_{+-} \\ \tilde{V}_{-+} & \tilde{V}_{--}} = U_0^{-1} (y) \bmat{U_{++} & U_{+-} \\ U_{-+} & U_{--}} + \bmat{4 m_1^2 / w^2 y & 0 \\ 0 & 4 m_1^2 / w^2 y} \;.
\eeq  The expression for the denominator is \beq
U_0 (y) = - 4 (2 J + 1) y^{5/2} (y+1)^{5/2} \alpha^2 \;,
\eeq and the expressions for the numerators are \bes \begin{align} \nonumber
	U_{++} = & \sqrt{y} \sqrt{y+1} \Bigl\lbrace 4 \left( 2 J+1 \right)  \left( y+1 \right)^2 \left( 2 y+1 \right)  \alpha^2 -4 J \left[ 2 J^2 +3 \left( J+1\right) \right]  y^2  \\ & -4 \left( J+1\right) \left[ 2 J \left( 2 J+1 \right) -1 \right]  y   -4 J^2 \left( 2 J+3 \right) -3 J+1 \Bigr\rbrace +4 J \left( J+1 \right) \, \left( y+1 \right) \left( 4 y^2 +3 y+1 \right) \;,
	\\
	U_{+-} = & \sqrt{J} \sqrt{J+1} \left\{ \sqrt{y} \sqrt{y+1} \left[ 4 y \left( 2 y+1 \right) +1 \right] -4 \left( y+1 \right) \left[ 2 y^2+ \left( J+1 \right) \left( 3 y+1 \right) \right] \right\} \;,
	\\
	U_{-+} = & \sqrt{J} \sqrt{J+1} \left\{ \sqrt{y} \sqrt{y+1} \left[ 4 y \left( 2 y+1 \right) +1 \right] -4 \left( y+1 \right) \left[ 2 y^2 -J \left( 3 y+1 \right) \right] \right\} \;,
	\\ \nonumber
	U_{--} = & \sqrt{y} \sqrt{y+1} \Bigl\lbrace 4 \left( 2 J+1 \right) \left( y+1 \right)^2 \left( 2 y+1 \right) \alpha^2 -4 \left( J+1 \right) \left[ J \left( 2 J+1 \right) +2 \right] y^2  \\ & -4 J \left[ 2 J \left( 2 J+3 \right) +1 \right]  y   -4 J^2 \left( 2 J+3 \right) -3 J \Bigr\rbrace -4 J \left( J+1 \right) \left( y+1 \right) \left( 4 y^2 +3 y+1 \right) \;.
\end{align} \ees  The final coupled equation in matrix form is thus \beq
\bmat{\tilde{H}_{++} & \tilde{H}_{+-} \\ \tilde{H}_{-+} & \tilde{H}_{--}} \bmat{u_+ \\ u_-} = d \bmat{u_+ \\ u_-} \;.
\eeq

\section{\label{sec:improvement}Iterative Improvement}
In each of the equations above, the final term of the potential $\tilde{V}$ (except for the cross terms in the coupled equation) is $4 m_1^2 / w^2 y$ left behind from our expansion of $\epsilon_w$; no other explicit factors of $w$ are present on the LHS.  The value of $w$ will in general be different for every state, thus as written our equations pertain to only a single state vector.  As a first approximation, we write $w \approx M = 2 m_1$ and replace the "mass term" with $1/y$, i.e. \beq
\tilde{H} \rightarrow \widehat{H} \equiv  \tilde{H} \left( \dfrac{4 m_1^2}{w^2 y} \rightarrow \dfrac{1}{y} \right) \;.
\eeq  With that substitution, we can solve for multiple radial eigenstates simultaneously, $\widehat{\msf{H}} \msf{V} = \msf{V} \msf{D}$, from which we get our first set of estimates $\widehat{w}_n$, then for each state $n$ we reverse the substitution \beq
\widehat{H} \rightarrow \tilde{H}_n \equiv \widehat{H} \left( \dfrac{1}{y} \rightarrow \dfrac{4 m_1^2}{\widehat{w}_n^2 y} \right) \;,
\eeq and solve $\tilde{\msf{H}}_n \mbf{V}_n = d_n \mbf{V}_n$.  A second iteration updating the value of $\widehat{w}_n$ is sufficient to converge the eigenvalue.

For the coupled equations, we begin the estimation procedure by focusing on the dominant portion of the Hamiltonian, $\widehat{H}_{++}$ for states $\prescript{3}{}L_{L+1}$ and  $\widehat{H}_{--}$ for states $\prescript{3}{}L_{L-1}$.  With estimates $\widehat{w}_n$ in hand, we then perform the iterative improvement on the full Hamiltonian $\tilde{\msf{H}}_n$, thereby including the contribution from the coupling terms.  The ARPACK library accepts as input suggestions for the starting value and vector, which we use to keep the algorithm from drifting during iteration.  Note that for a particular state $n$ the same value of $\widehat{w}_n$ is used in both $\tilde{H}_{++}$ and $\tilde{H}_{--}$ and that either $\norm{u_+} \gg \norm{u_-}$ or vice versa.

\section{\label{sec:perturbative}Perturbative Results}
The point is made in~\cite{crater1992PhysRevD.46.5117} that one should verify the results of one's nonperturbative solution method by confirming its agreement with perturbative results in an appropriate regime before trusting its results in a regime where perturbative methods fail (for example by varying the strength of the coupling $\alpha$).  Those authors present in Eqns.(5.18) through (5.25) a series of analytic formulae for the perturbative solution of the spectrum.  They also present in Table~1 therein numerical values from their evaluation of the formulae.  When we evaluate their formulae as published, we do not get the same result.  Two numerical factors are present, the coupling strength $\alpha = 1 / 137.0359895$ which they published, and the unit conversion factor $m_e c^2$ which they did not publish; we are using value $m_e c^2 = 0.51099895069$ MeV.  We therefor expect that our numbers should agree with theirs for all states up to a constant numerical factor from the conversion of units.  That is not what we found.

\begin{table}
	\centering
	\begin{tabular}{|cccc|l|l|l|l|l|}
		\hline
		$L$ & $S$ & $J$ & $n$ & \multicolumn{1}{c|}{$P_1$} & \multicolumn{1}{c|}{$P_2$} & \multicolumn{1}{c|}{$P_{92}$} & \multicolumn{1}{c|}{$P_1 / P_{92}$} & \multicolumn{1}{c|}{$P_2 / P_{92}$} \\
		\hline
		0 & 0 & 0 & 1 & -6.803322989669 & -6.803322989669 & -6.803325627900 &  0.999999612215 &  0.999999612215 \\ 
		0 & 1 & 1 & 1 & -6.802839975141 & -6.802839975141 & -6.802842613200 &  0.999999612212 &  0.999999612212 \\ 
		0 & 0 & 0 & 2 & -1.700786879887 & -1.700786879887 & -1.700787539400 &  0.999999612231 &  0.999999612231 \\ 
		0 & 1 & 1 & 2 & -1.700726503071 & -1.700726503071 & -1.700727162600 &  0.999999612208 &  0.999999612208 \\ 
		1 & 1 & 0 & 2 & -1.700756691479 & -1.700756691479 & -1.700757351000 &  0.999999612219 &  0.999999612219 \\ 
		1 & 0 & 1 & 2 & -1.700756691479 & -1.700726503071 & -1.700727162600 &  1.000017362502 &  0.999999612208 \\ 
		1 & 1 & 1 & 2 & -1.700771785683 & -1.700734050173 & -1.700734709700 &  1.000021799980 &  0.999999612211 \\ 
		1 & 1 & 2 & 2 & -1.700715937128 & -1.700715937128 & -1.700716596600 &  0.999999612239 &  0.999999612239 \\ 
		0 & 0 & 0 & 3 & -0.755895706311 & -0.755895706311 & -0.755895999400 &  0.999999612263 &  0.999999612263 \\ 
		0 & 1 & 1 & 3 & -0.755877816884 & -0.755877816884 & -0.755878110000 &  0.999999612218 &  0.999999612218 \\ 
		1 & 1 & 0 & 3 & -0.755886761598 & -0.755886761598 & -0.755887054700 &  0.999999612241 &  0.999999612241 \\ 
		1 & 0 & 1 & 3 & -0.755886761598 & -0.755877816884 & -0.755878110000 &  1.000011445758 &  0.999999612218 \\ 
		1 & 1 & 1 & 3 & -0.755891233954 & -0.755880053063 & -0.755880346200 &  1.000014404071 &  0.999999612191 \\ 
		1 & 1 & 2 & 3 & -0.755874686234 & -0.755874686234 & -0.755874979300 &  0.999999612283 &  0.999999612283 \\ 
		2 & 1 & 1 & 3 & -0.755876475177 & -0.755876475177 & -0.755876768300 &  0.999999612208 &  0.999999612208 \\ 
		2 & 0 & 2 & 3 & -0.755879605827 & -0.755874238999 & -0.755874532100 &  1.000006712393 &  0.999999612236 \\ 
		2 & 1 & 2 & 3 & -0.755880500298 & -0.755874686234 & -0.755874979300 &  1.000007304116 &  0.999999612283 \\ 
		2 & 1 & 3 & 3 & -0.755872961183 & -0.755872961183 & -0.755873254300 &  0.999999612213 &  0.999999612213 \\ 
		\hline
	\end{tabular}
	\caption{Comparison of perturbative results for binding energy in eV}
	\label{tab:comppert} 
\end{table}

\begin{table}
	\centering
	\begin{tabular}{|cccc|l|l|l|}
		\hline
		$L$ & $S$ & $J$ & $n$ & \multicolumn{1}{c|}{$\eta_1$} & \multicolumn{1}{c|}{$\eta_2$} & \multicolumn{1}{c|}{$\eta_{QED}$} \\
		\hline
0 & 0 & 0 & 1 & -2.000000 & -2.000000 & -2 \\ 
0 & 1 & 1 & 1 & -0.666667 & -0.666667 & -2/3 \\ 
1 & 1 & 0 & 2 & -1.333333 & -1.333333 & -4/3 \\ 
1 & 0 & 1 & 2 & -1.333333 & -0.666667 & -2/3 \\ 
1 & 1 & 1 & 2 & -1.666667 & -0.833333 & -5/6 \\ 
		\hline
	\end{tabular}
	\caption{Comparison of perturbative coefficient $\eta_{\pm, \kappa}$}
	\label{tab:compperteta} 
\end{table}

What we found is that \revtext{the evaluation of} the formula for the case $J = L >= 1$ requires an additional factor of $1/2$ applied to $\eta_\pm$ to come into agreement \revtext{with the numbers published in Table~1 of~\cite{crater1992PhysRevD.46.5117}}.  Explicitly, when we use \revtext{their} expression \beq \label{eqn:etano2}
\eta_\pm = \left[ a + c \pm \sqrt{(a-c)^2 + 4 b^2} \right] \;,
\eeq in terms of \begin{align}
	a & = -2 / (2 L + 1) \;, \\
	b & = (m_2 - m_1) / [M (2 L + 1) \sqrt{L (L + 1)} ] \;, \\
	c & = a - 1 / [ (2 L + 1) L (L + 1) ] \;,
\end{align} we get the values under column \revtext{$P_1$} in Table~\ref{tab:comppert}.  \revtext{If instead we use the expression $\eta_\pm \rightarrow \eta_\pm / 2$}, i.e. \beq \label{eqn:etaw2}
\eta_\pm = \left[ a + c \pm \sqrt{(a-c)^2 + 4 b^2} \right] / 2 \;,
\eeq we get the values under column \revtext{$P_2$}.  For all other cases, $J \neq L$ or $J = L = 0$, we agree on the value of $\eta_\kappa$ found in their Eqns.~(5.19) and (5.20).  In all cases the binding energy can then be written as \beq
w - M = - \dfrac{\mu \alpha^2}{2 n^2} + \dfrac{\mu \alpha^4}{8 n^4} \left( 3 - \dfrac{\mu}{M} \right) + \dfrac{\mu \alpha^4}{2 n^3} \eta_{\pm, \kappa} \;,
\eeq where the reduced and constituent masses are $\mu = m_1 / 2$ and $M = 2 m_1$ and $n$ is the total (orbital plus radial) quantum number.

Under column $P_{92}$ in Table~\ref{tab:comppert} we show the values from Table~1 in~\cite{crater1992PhysRevD.46.5117}.  In the last two columns we show the ratios $P_1 / P_{92}$ and $P_2 / P_{92}$.  For the case of $\eta_\kappa$ we find values for the ratio $< 1$ constant down to 10 decimal places, after which the fluctuation is attributable to the truncation of the values under $P_{92}$ to that precision.  However, for the case of $\eta_\pm$, values for $P_1 / P_{92}$ are $> 1$ and fluctuate at the 6th decimal place.  Playing around with $+$ and $-$ signs (common misprints) did not help, but playing around with a factor of 2 \revtext{brought the evaluation of the binding energy into agreement with the published table and is supported by the numerical results.}
The authors~\cite{crater1992PhysRevD.46.5117} in Endnote [50] therein discuss the difficulties investigators have had getting the correct formula for the positronium spectrum~\cite{bethesalpeter1957oneandtwo,connell1991PhysRevD.43.1393}.  \revtext{In Table~\ref{tab:compperteta} we compare our calculated values for $\eta_{\pm, \kappa}$ using Eqn.~\ref{eqn:etano2} for $\eta_1$ and Eqn.~(\ref{eqn:etaw2}) for $\eta_2$.  In the final column is the commonly accepted value $\eta_{QED}$ as evaluated by~\cite{karplus1952PhesRev.87.848,fulton1954PhysRev.95.811}.  Interestingly, the value for $\eta_k$ for state $\prescript{3}{}S_1$ is in agreement with $\eta_{QED}$ despite the absence of the effect of virtual photon annihilation in the quasipotential formulation.  Nonetheless, the expected degeneracy when annihilation is neglected between states $2\prescript{3}{}S_1$ and $1\prescript{1}{}P_1$ when $n = 2$ is observed in columns $P_2$ and $P_{92}$ in Table~\ref{tab:comppert}.  See Endnote~55 in~\cite{crateretalORNL12122}.}

\section{\label{sec:nonperturbative}Nonperturbative Results}
We are close to being able to present our nonperturbative results for the positronium spectrum, but first we must dispense with some numerical discussion.  Our standard axes $\tilde{x}$, $\tilde{y}$, and $\tilde{z}$ each have $N = 5000$ points linearly spaced.  Since $\tilde{y}_N = 1$, then $\tilde{y}_1 = 1/N = 2\textsc{e}-4$.  Instead of $\log \tilde{y}_1 \sim -8.5$ as the lower limit of $\tilde{z}$ we use a range $\tilde{z} \in [-25, 0]$ that probes deeper towards the origin $r = 0$.  Our stencil for finite differences is of order 3 thus length 7, i.e. spans points $[ -3, -2, -1, 0, 1, 2, 3 ]$, and edge correction is applied to $\tilde{x}$ and $\tilde{y}$ coordinates.  In Table~\ref{tab:scalefactors} we show the scale factors used in this numerical analysis, conditioned on the value $L$ of the state.

\begin{table}
	\centering
	\begin{tabular}{|c|c|c|c|}
		\hline
		$L$ & $s_x$ & $s_y$ & $s_z$ \\
		\hline
		0 & 1.3\textsc{e}4 & 1.7\textsc{e}6 & $\exp 15$ \\
		1 & 6.0\textsc{e}4 & 2.8\textsc{e}6 & $\exp 20$ \\
		$> 1$ & 9.9\textsc{e}4 & 4.6\textsc{e}6 & $\exp 20$ \\
		\hline
		\hline
	\end{tabular}
	\caption{Scale factors used in numerical analysis}
	\label{tab:scalefactors} 
\end{table}

\begin{table}
	\centering
	\begin{tabular}{|ccccc|r|r|r|r|r|r|}
		\hline
		$J$ & $S$ & $L$ & $R$ & $n$ & \multicolumn{1}{c|}{$x$} & \multicolumn{1}{c|}{$y$} & \multicolumn{1}{c|}{$z$} & \multicolumn{1}{c|}{$\chi^{10}_x$} & \multicolumn{1}{c|}{$\chi^{10}_y$} & \multicolumn{1}{c|}{$\chi^{10}_z$} \\
		\hline
0 & 0 & 0 & 1 & 1 &  -6.803323011011 &  -6.803318891907 &  -6.803323010784 & -13 & -16 &  -7 \\ 
1 & 1 & 0 & 1 & 1 &  -6.802986183749 &  -6.803314471328 &  -6.802840028347 & -13 & -16 &  -9 \\ 
0 & 0 & 0 & 2 & 2 &  -1.700786882787 &  -1.700786368112 &  -1.700786882787 & -13 & -16 &  -7 \\ 
0 & 1 & 1 & 1 & 2 &  -1.700756693952 &  -1.700756693952 &  -1.700756693952 & -15 & -16 & -14 \\ 
1 & 1 & 1 & 1 & 2 &  -1.700734050624 &  -1.700734050624 &  -1.700734050397 & -15 & -16 & -15 \\ 
1 & 0 & 1 & 1 & 2 &  -1.700726503189 &  -1.700726503415 &  -1.700726503189 & -15 & -16 & -15 \\ 
1 & 1 & 0 & 2 & 2 &  -1.700744779833 &  -1.700785815766 &  -1.700726509769 & -13 & -16 &  -9 \\ 
2 & 1 & 1 & 1 & 2 &  -1.700715937255 &  -1.700715937255 &  -1.700715937255 & -15 & -16 & -15 \\ 
0 & 0 & 0 & 3 & 3 &  -0.755895707061 &  -0.755894273323 &  -0.755895707061 & -13 & -16 &  -7 \\ 
0 & 1 & 1 & 2 & 3 &  -0.755886762423 &  -0.755886762423 &  -0.755886762423 & -15 & -16 & -14 \\ 
1 & 1 & 1 & 2 & 3 &  -0.755880053264 &  -0.755880053264 &  -0.755880053264 & -15 & -16 & -15 \\ 
1 & 0 & 1 & 2 & 3 &  -0.755877816877 &  -0.755877817104 &  -0.755877816877 & -15 & -16 & -15 \\ 
1 & 1 & 0 & 3 & 3 &  -0.755883232314 &  -0.755894109480 &  -0.755877818919 & -13 & -16 &  -9 \\ 
1 & 1 & 2 & 1 & 3 &  -0.755876475272 &  -0.755876475272 &  -0.755876475272 & -15 & -16 & -16 \\ 
2 & 1 & 1 & 2 & 3 &  -0.755874686163 &  -0.755874686390 &  -0.755874686163 & -15 & -16 & -15 \\ 
2 & 1 & 2 & 1 & 3 &  -0.755874686163 &  -0.755874686163 &  -0.755874686163 & -15 & -16 & -16 \\ 
2 & 0 & 2 & 1 & 3 &  -0.755874239113 &  -0.755874239113 &  -0.755874239113 & -15 & -16 & -16 \\ 
3 & 1 & 2 & 1 & 3 &  -0.755872961275 &  -0.755872961275 &  -0.755872961275 & -15 & -16 & -16 \\ 
		\hline
	\end{tabular}
	\caption{Nonperturbative estimates of the binding energy in eV for positronium}
	\label{tab:nonpertbinding} 
\end{table}

We can now show in Table~\ref{tab:nonpertbinding} the results of our nonperturbative solution for the excited states of the positronium spectrum.  For each state the estimate of the binding energy $w - M$ in units of eV is displayed under the heading for the coordinate axis employed.  In the final three columns we show a value which is the $\log_{10}$ of the norm of the residual vector, evaluated for each state according to \beq
\chi^{10}_c \equiv \log_{10} \norm{\msf{H}_k \mbf{V}_k - d_k \mbf{V}_k} \;,
\eeq for coordinate $c$ and state identified by $k$ using the final values from the iterative improvement, which serves as an estimate of the \revtext{precision of our solution of the eigensystem}.  We like to include explicitly in the list of quantum numbers the radial number $R$ (implicit in previous tables) in terms of the number of absolute peaks (not nodes); that value needs to be checked through inspection of the eigenvectors during numerical analysis to be sure one is recovering the desired states.  We can see from the values of $\chi^{10}$ that coordinate $y$ tends to overestimate its confidence, especially for states with $L = 0$, compared to $x$ and $z$ estimates which we feel are more robust.  Ideally, one would see numerical agreement across all three coordinate estimates, and for the majority of states that is indeed what we see.  \revtext{Note that for the states associated with $\eta_\pm$, e.g. $(J, S, L) = (1, 0, 1)$, the numerical results clearly favor $P_2$ over $P_1$ as the correct evaluation of the analytic perturbative expression.}

\begin{figure}
	\centering
	\includegraphics[width=\textwidth]{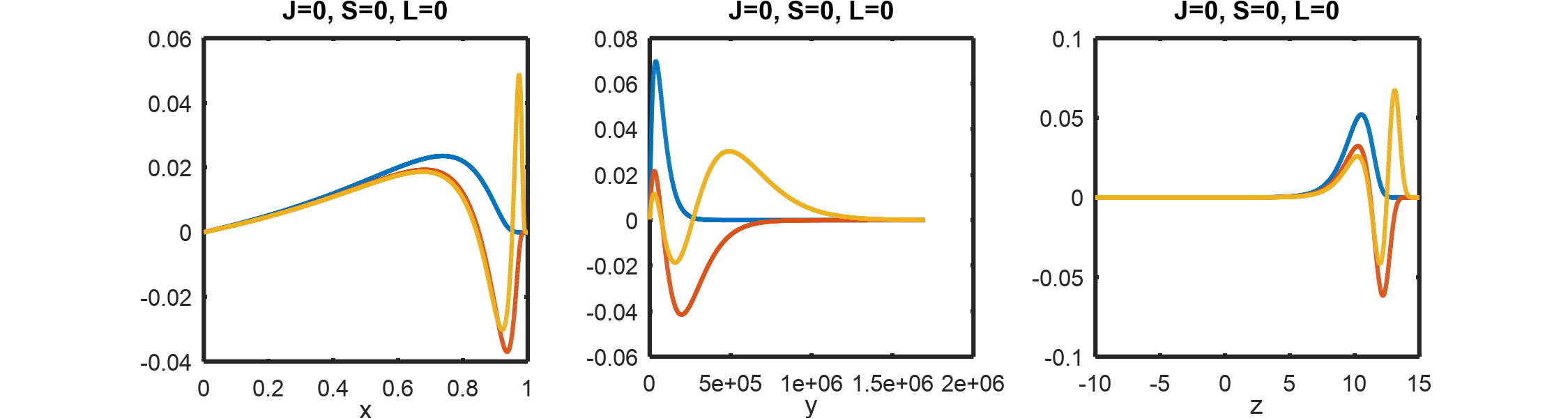}
	\caption{Eigenvectors for state $\prescript{1}{}S_0$ over coordinates $x$, $y$, and $z$}
	\label{fig:B_000}
\end{figure}

\begin{figure}
	\centering
	\includegraphics[width=\textwidth]{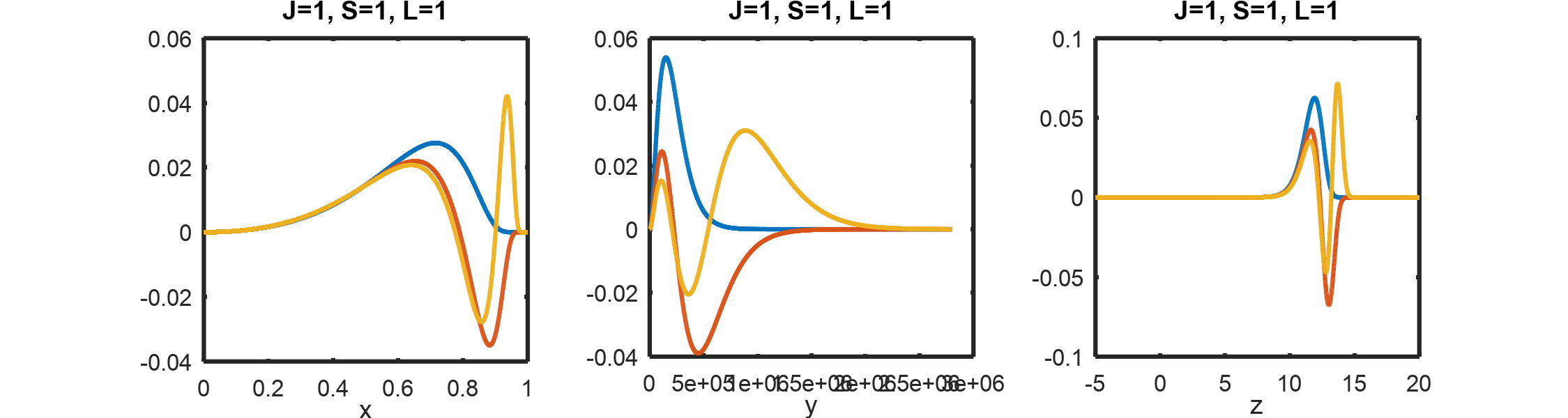}
	\caption{Eigenvectors for state $\prescript{3}{}P_1$ over coordinates $x$, $y$, and $z$}
	\label{fig:B_111}
\end{figure}

\begin{figure}
	\centering
	\includegraphics[width=\textwidth]{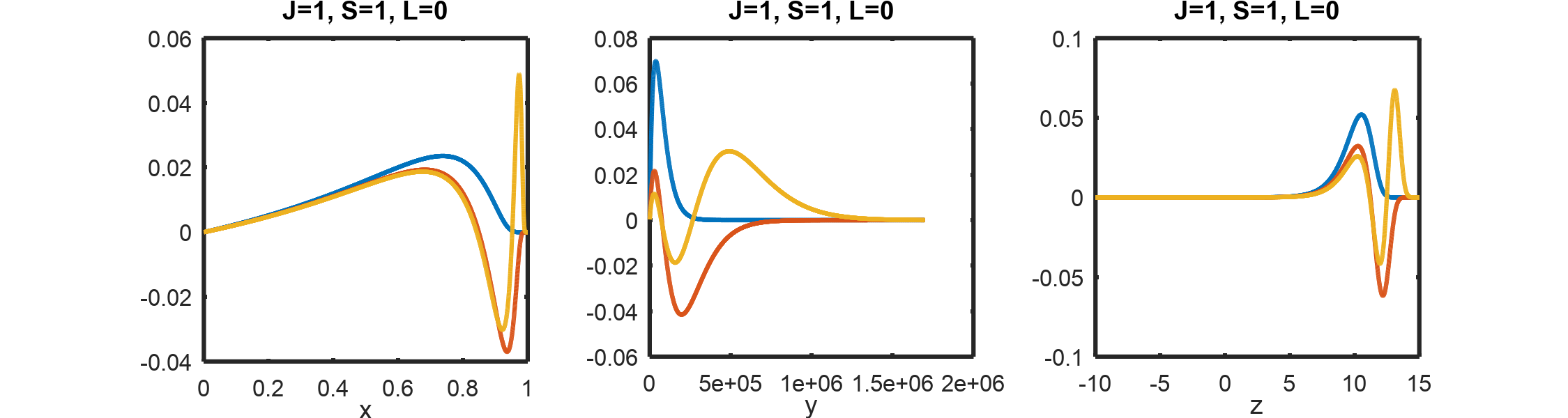}
	\caption{Eigenvectors for state $\prescript{3}{}S_1$ over coordinates $x$, $y$, and $z$}
	\label{fig:B_110}
\end{figure}

\begin{figure}
	\centering
	\includegraphics[width=\textwidth]{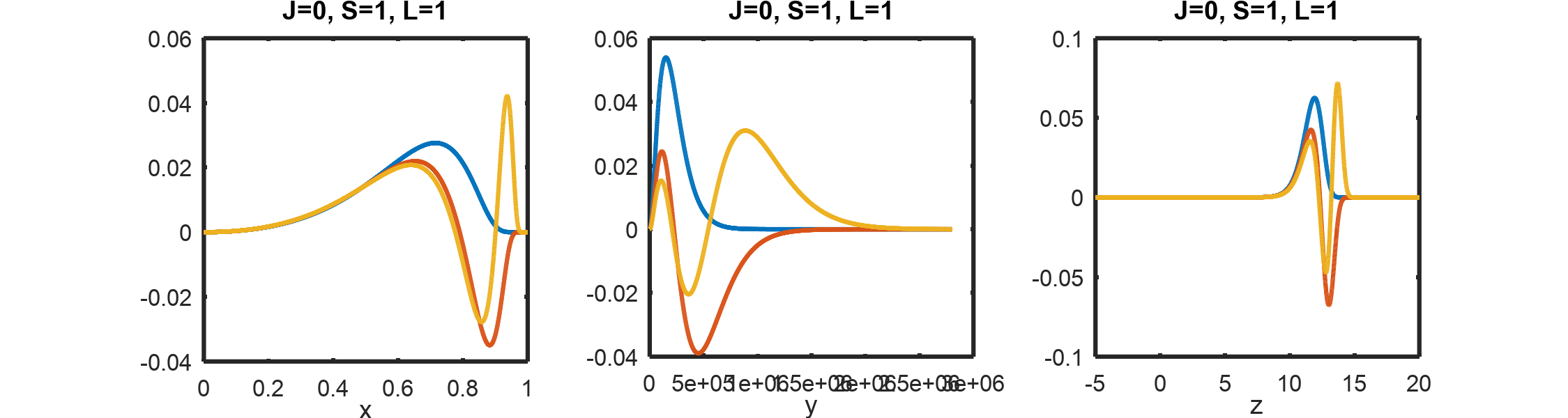}
	\caption{Eigenvectors for state $\prescript{3}{}P_0$ over coordinates $x$, $y$, and $z$}
	\label{fig:B_011}
\end{figure}

It is instructive to view the eigenvectors obtained for the various states.  If the suggested starting value and vector are far from ideal, the numerical algorithm can return states other than those desired.  In this work we verify the radial quantum number by calculating the number of absolute peaks in the state vector.  In Figs.~\ref{fig:B_000}, \ref{fig:B_111}, \ref{fig:B_110}, and \ref{fig:B_011} we display the eigenvectors for the first three radial states in each channel, $\prescript{1}{}J_J$, $\prescript{3}{}J_J$, $\prescript{3}{}L_{L+1}$, and $\prescript{3}{}L_{L-1}$, respectively.  For the states with $L = 0$, note how we use the scale factor $s_x$ to emphasize the portion of the eigenvector close to the origin, which was necessary to achieve agreement with the perturbative results.

\begin{table}
	\centering
	\begin{tabular}{|ccccc|r|r|r|}
		\hline
		$J$ & $S$ & $L$ & $R$ & $n$ & \multicolumn{1}{c|}{$\Delta^{10}_x$} & \multicolumn{1}{c|}{$\Delta^{10}_y$} & \multicolumn{1}{c|}{$\Delta^{10}_z$} \\
		\hline
0 & 0 & 0 & 1 & 1 & -8.5 & -6.2 & -8.5 \\ 
1 & 1 & 0 & 1 & 1 & -4.7 & -4.2 & -8.1 \\ 
0 & 0 & 0 & 2 & 2 & -8.8 & -6.5 & -8.8 \\ 
0 & 1 & 1 & 1 & 2 & -8.8 & -8.8 & -8.8 \\ 
1 & 1 & 1 & 1 & 2 & -9.6 & -9.6 & -9.9 \\ 
1 & 0 & 1 & 1 & 2 & -10.2 & -9.7 & -10.2 \\ 
1 & 1 & 0 & 2 & 2 & -5.0 & -4.5 & -8.4 \\ 
2 & 1 & 1 & 1 & 2 & -10.1 & -10.1 & -10.1 \\ 
0 & 0 & 0 & 3 & 3 & -9.0 & -5.7 & -9.0 \\ 
0 & 1 & 1 & 2 & 3 & -9.0 & -9.0 & -9.0 \\ 
1 & 1 & 1 & 2 & 3 & -9.6 & -9.6 & -9.6 \\ 
1 & 0 & 1 & 2 & 3 & -11.0 & -9.5 & -11.0 \\ 
1 & 1 & 0 & 3 & 3 & -5.1 & -4.7 & -8.6 \\ 
1 & 1 & 2 & 1 & 3 & -9.9 & -9.9 & -9.9 \\ 
2 & 1 & 1 & 2 & 3 & -10.0 & -9.7 & -10.0 \\ 
2 & 1 & 2 & 1 & 3 & -10.0 & -10.0 & -10.0 \\ 
2 & 0 & 2 & 1 & 3 & -9.8 & -9.8 & -9.8 \\ 
3 & 1 & 2 & 1 & 3 & -9.9 & -9.9 & -9.9 \\ 
		\hline
		&   &   &   &   & -8.8 & -8.2 & -9.4 \\
		\hline
	\end{tabular}
	\caption{Comparison of nonperturbative and perturbative estimates}
	\label{tab:nonpertcompare} 
\end{table}

\clearpage

\clearpage

Finally, let us compare our nonperturbative estimates for the binding energy to those obtained from perturbation theory, \revtext{which is an assessment of the accuracy of our solution to the eigensystem}.  For this comparison, we calculate a value $\Delta^{10}$ defined for each state and coordinate $c$ as the $\log_{10}$ of the absolute value of the difference of the ratio of the estimates from unity, i.e. \beq
\Delta^{10}_c = \log_{10} \abs{1 - \epsilon_c / P_2} \;,
\eeq where $\epsilon_c$ is the nonperturbative estimate for the binding energy in Table~\ref{tab:nonpertbinding} and $P_2$ is the perturbative estimate from Table~\ref{tab:comppert}.  The value of $\Delta^{10}_c$ is roughly a measure of the number of decimal digits of agreement between our estimates, and we show in Table~\ref{tab:nonpertcompare} the result of our comparison.  In the final row we show a value averaged over all states.  From this comparison we see that coordinate $z$ provides the best nonperturbative estimate with coordinate $x$ not far behind; it is coordinate $y \propto r$ which is least accurate for numerical evaluation.  It would be nice if we could achieve the level of agreement found by~\cite{crater1992PhysRevD.46.5117}; however, while they mention in~\cite{crateretalORNL12122} some details of their numerical procedure, the ``separate publication'' with ``details of that procedure'' has never appeared.  Their description of inverting a large banded matrix sounds sufficiently close to what we are doing with ARPACK that we believe our results are in agreement with theirs.

\section{\label{sec:comparison}Comparison To Improved Breit Equation}
The same physics can be cast into a form~\cite{mourad1995covariantbreit,crateretal1996singularityfree,vanalstine1997threeequations} that resembles the traditional Breit equation~\cite{breit1932PhysRev.39.616}.  Again, the purpose of this article is not to expound upon the model's derivation but rather to investigate its application.  The essential difference from the original Breit equation is the elimination of unphysical degrees of freedom in the center of mass frame of reference.

The main result of Ref.~\cite{crateretal1996singularityfree} is the decomposition of the model equations into two sectors of opposite parity $\mcal{P} = (-1)^{J+1}$ or $(-1)^J$.  We will restrict consideration to $J = 0$ and equal mass systems $m_2 = m_1$.  The 16 component spinor is written in terms of singlet and triplet wave functions \beq \label{eqn:psi16}
\msf{\Psi} = \bmat{\psi, \phi, i \chi, \eta \\ \mbs{\psi}, \mbs{\phi}, i \mbs{\chi}, \mbs{\eta}} \;,
\eeq making use of vector spherical harmonics $\mbf{X}$, $\mbf{Y}$, and $\mbf{Z}$, and each sector comprises 4 algebraic and 4 differential equations coupling 8 components of $\msf{\Psi}$.  For $J = 0$ only the harmonic $\mbf{Y}$ proportional to the displacement vector $\mbf{r}$ is considered, which eliminates 4 components from each sector.  The remaining algebraic equations can be solved formally, leaving 2 coupled first order differential equations in each sector.

\begin{figure}
	\centering
	\includegraphics[width=\textwidth]{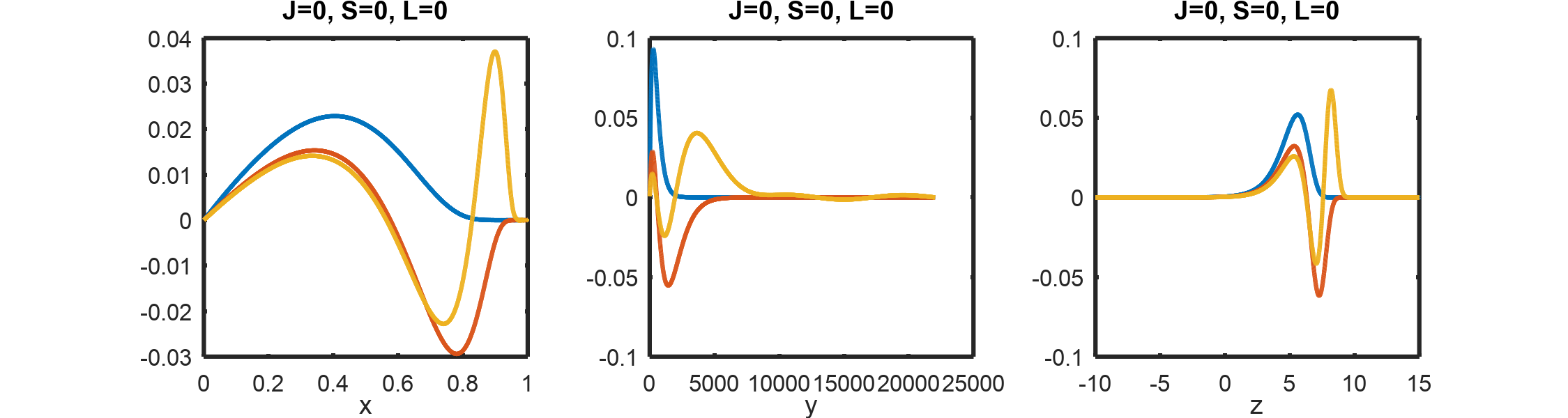}
	\caption{Eigenvectors for state $\prescript{1}{}S_0$ over coordinates $x$, $y$, and $z$ from Eqn.~(\ref{eqn:w1})}
	\label{fig:C_000}
\end{figure}

For $\mcal{P} = -$ the surviving components are $\phi(r)$ and $\psi_Y(r)$ with equation \beq
\bmat{\dfrac{4 m_1^2 / w}{(1+2 \alpha / w r)} & \dfrac{4/r+2\,d/dr}{(1+2 \alpha / w r)^2} \\ -2\,d/dr & 0} \bmat{\phi \\ \psi_Y} = w \bmat{\phi \\ \psi_Y} \;,
\eeq where $\phi \gg \psi_Y$ is the dominant singlet component.  Substituting $\psi_Y(r) = -(2/w) d\phi(r)/dr$ into the upper equation yields the second order equation \beq \label{eqn:w1}
\left[ - \dfrac{4 \alpha^2}{w r^2} - \dfrac{4 \alpha}{r} + \dfrac{4 m_1^2}{w} \left( 1 + \dfrac{2 \alpha}{w r} \right) - \dfrac{4}{w} \dfrac{d^2}{dr^2} \right] \phi(r) = w \phi(r) \;,
\eeq where we have absorbed a factor $\phi(r) \rightarrow \phi(r)/r$; the same equation obtains if that order of operations is reversed. \revtext{If we then attempt to solve that equation naively using the techniques of the preceding section,} the solution we get is not what we expect.  In Table~\ref{tab:bindingw1} we show the numerical results obtained, \revtext{and in Figure~\ref{fig:C_000} we show the corresponding eigenvectors.  Notice that the binding energies are off by a factor of 2 and that the third radial state has not converged properly.  These results are curious and bear some further investigation.}

\revtext{
If we move a factor of $w$ from LHS to RHS, \beq \label{eqn:w2}
\left[ - \dfrac{4 \alpha^2}{r^2} - \dfrac{4 \alpha w}{r} + 4 m_1^2 \left( 1 + \dfrac{2 \alpha}{w r} \right) - 4 \dfrac{d^2}{dr^2} \right] \phi(r) = w^2 \phi(r) \;,
\eeq our equation is equivalent to Eqn.~(\ref{eqn:JLS0}) multiplied by 4, after substituting for $J$, $\epsilon_w$, and $b^2_w$, and it can be solved for the physically correct spectrum.  Formally, dividing Eqn.~(\ref{eqn:w2}) by a factor of $w$ is equivalent to taking the square root of the operator on the LHS.  In other words, that algebraic manipulation amounts to attempting to linearize our relativistic Hamiltonian operator.  Let the true relativistic energy be notated $w_\Omega = 2 m_1 - E_b$, where $E_b$ is the true binding energy.  When we iterate our solution $w_k$ for Eqn.~(\ref{eqn:w1}), we are asking ARPACK to solve $H(w_k) / w_k$ for an eigenvalue $d_{k+1} = w_\Omega^2 / w_k$.  Since $w_k$ is close to the constituent energy $2 m_1$, we can approximate \beq
d_{k+1} \approx \dfrac{(2 m_1 - E_b)^2}{2 m_1} \approx 2 m_1 - 2 E_b \;,
\eeq and we see that the distance from the constituent energy $2 m_1$ to the eigenvalue $d_{k+1}$ is exactly $2 E_b$.
}

\begin{table}
	\centering
	\begin{tabular}{|c|r|r|r|r|r|r|}
		\hline
		$n$ & \multicolumn{1}{c|}{$x$} & \multicolumn{1}{c|}{$y$} & \multicolumn{1}{c|}{$z$} & \multicolumn{1}{c|}{$\chi^{10}_x$} & \multicolumn{1}{c|}{$\chi^{10}_y$} & \multicolumn{1}{c|}{$\chi^{10}_z$} \\
		\hline
1 & -13.610246616623 & -13.612396498970 & -13.611357269008 & -8.2 & -7.3 & -7.7 \\ 
2 & -3.402288305918 & -3.401942103533 & -3.401062853835 & -9.7 & -8.9 & -8.8 \\ 
3 &  0.000000000000 &  0.000000000000 &  0.000052474863 & -Inf & -Inf & -5.5 \\ 
		\hline
	\end{tabular}
	\caption{Nonperturbative estimates of the binding energy in eV for state $\prescript{1}{}S_0$ from Eqn.~(\ref{eqn:w1})}
	\label{tab:bindingw1} 
\end{table}

\revtext{
Either Eqn.~(\ref{eqn:w1} or Eqn.~(\ref{eqn:w2}) can be cast into the equivalent form \beq \label{eqn:w3}
\left[ - \dfrac{4 \alpha^2 w}{r^2} - \dfrac{4 \alpha w^2}{r} + 4 m_1^2 w +  \dfrac{8 m_1^2 \alpha}{r} - 4 w \dfrac{d^2}{dr^2} \right] \phi(r) = w^3 \phi(r) \;,
\eeq which can be put into standard cubic form $(A_3 w^3 + A_2 w^2 + A_1 w_1 + A_0) \phi(r) = 0$ using \beq
	A_3 = -I \;, A_2 = - \dfrac{4 \alpha}{r} \;, A_1 = - \dfrac{4 \alpha^2}{r^2} + 4 m_1^2 I - 4 \dfrac{d^2}{dr^2} \;, A_0 = \dfrac{8 m_1^2 \alpha}{r} \;,
\eeq where $I$ is the identity operator.  Let us also define the auxiliary vector $\Psi = [\psi_0, \psi_1, \psi_2]$, where $\psi_j(r) \equiv w^j \phi(r)$, and make use of the relation \beq
w^3 \phi = w \psi_2 = \sum_{j=0}^2 A_j \psi_j
\eeq to write the generalized eigenvalue problem $\msf{A} \Psi = w \msf{B} \Psi$ in block matrix form \beq \label{eqn:blockA}
\bmat{0 & I & 0 \\ 0 & 0 & I \\ A_0 & A_1 & A_2} \Psi = w \bmat{I & 0 & 0 \\ 0 & I & 0 \\ 0 & 0 & I} \Psi \;,
\eeq which reduces to a standard eigenvalue problem $\msf{A} \Psi = w \Psi$ since $\msf{B}$ is just the identity operator.  That problem can be solved by ARPACK for a set of radial states simultaneously since all factors of $w$ are on the RHS, but the states beyond the lowest benefit from an iteration of improvement to refine the eigenvalue and eigenvector.  The original wavefunction $\phi(r)$ can be extracted from the uppermost component of $\Psi$.  In Table~\ref{tab:bindingw3} we show the numerical results obtained from solving Eqn.~(\ref{eqn:w3}).
}

\begin{table}
	\centering
	\begin{tabular}{|c|r|r|r|r|r|r|}
		\hline
		$n$ & \multicolumn{1}{c|}{$x$} & \multicolumn{1}{c|}{$y$} & \multicolumn{1}{c|}{$z$} & \multicolumn{1}{c|}{$\chi^{10}_x$} & \multicolumn{1}{c|}{$\chi^{10}_y$} & \multicolumn{1}{c|}{$\chi^{10}_z$} \\
		\hline
1 & -6.803322908553 & -6.803315676095 & -6.803318549131 & -13.4 & -15.3 & -8.7 \\ 
2 & -1.700786869965 & -1.700785965993 & -1.700786325109 & -13.5 & -15.3 & -8.8 \\ 
3 & -0.755895703317 & -0.755895435314 & -0.755895541743 & -13.5 & -15.3 & -8.9 \\ 
		\hline
	\end{tabular}
	\caption{Nonperturbative estimates of the binding energy in eV for state $\prescript{1}{}S_0$ from Eqn.~(\ref{eqn:w3})}
	\label{tab:bindingw3} 
\end{table}

For $\mcal{P} = +$ the surviving components are $\psi(r)$ and $\phi_Y(r)$ with equation \beq
\bmat{0 & \dfrac{4/r+2\,d/dr}{(1+2 \alpha / w r)^2} \\ -2\,d/dr & \dfrac{4 m_1^2 / w}{(1+2 \alpha / w r)}} \bmat{\psi \\ \phi_Y} = w \bmat{\psi \\ \phi_Y} \;.
\eeq  This time the dominant component $\phi_Y \gg \psi$ is a member of the triplet representation, thus we expect to recover the binding energies for the $\prescript{3}{}P_0$ state.  \revtext{After considerable manipulation (see breitJ0pos.wxmx among the Supplemental Material), one can derive the second order equation \beq \label{eqn:breit3P0}
\left[ \dfrac{m_1^2 (2 \alpha + r w)}{r w} - \dfrac{(2 \alpha + r w)^2}{4 r^2} + \dfrac{2 w}{r (2 \alpha + r w)} - \dfrac{4 \alpha}{r (2 \alpha + r w)} \dfrac{d}{dr} - \dfrac{d^2}{dr^2} \right] \phi(r) = 0 \;.
\eeq  The equivalently scaled form of Eqn.~(\ref{eqn:3P0}) reads \beq \label{eqn:tbde3P0}
\left[ \dfrac{m_1^2 (2 \alpha + r w)}{r w} - \dfrac{(2 \alpha + r w)^2}{4 r^2} + \dfrac{2 w^2}{(2 \alpha + r w)^2} - \dfrac{d^2}{dr^2} \right] \phi(r) = 0 \;,
\eeq and we can see immediately that the form of the quasipotential term is not the same.}  Nevertheless, when we solve Eqn.~(\ref{eqn:breit3P0}) using ARPACK, we recover exactly the same spectrum as produced by Eqn.~(\ref{eqn:tbde3P0}), as shown in Table~\ref{tab:comp3P0}, leaving us to suspect that an additional transformation would bring the equations into formal agreement.  Note that some transformations have occurred in going from ~\cite{crater1992PhysRevD.46.5117} upon which ~\cite{crateretal1996singularityfree} is based to ~\cite{crater2012PhysRevD.85.116005} from which Eqn.~(\ref{eqn:3P0}) is drawn.

\begin{table}
	\centering
	\begin{tabular}{|c|r|r|r|r|r|r|}
		\hline
		$n$ & \multicolumn{1}{c|}{$x$} & \multicolumn{1}{c|}{$y$} & \multicolumn{1}{c|}{$z$} & \multicolumn{1}{c|}{$\chi^{10}_x$} & \multicolumn{1}{c|}{$\chi^{10}_y$} & \multicolumn{1}{c|}{$\chi^{10}_z$} \\
		\hline
2 & -1.700756693952 & -1.700756693952 & -1.700756693952 & -15 & -15 & -13 \\ 
3 & -0.755886762423 & -0.755886762310 & -0.755886762310 & -15 & -15 & -13 \\ 
4 & -0.425184544673 & -0.425184535028 & -0.425184544673 & -15 & -15 & -13 \\ 
		\hline
2 & -1.700756693952 & -1.700756693952 & -1.700756693952 & -15 & -15 & -14 \\ 
3 & -0.755886762423 & -0.755886762310 & -0.755886762310 & -15 & -15 & -14 \\ 
4 & -0.425184544673 & -0.425184535028 & -0.425184544673 & -15 & -15 & -14 \\ 
		\hline
	\end{tabular}
	\caption{Nonperturbative estimates of the binding energy in eV for state $\prescript{3}{}P_0$ from Eqn.~(\ref{eqn:breit3P0}) compared to  Eqn.~(\ref{eqn:tbde3P0})}
	\label{tab:comp3P0} 
\end{table}

\section{\label{sec:conclusion}Conclusion}
In conclusion, we have estimated the excited states of positronium from the two body Dirac equations of constraint using both a perturbative formula and nonperturbative techniques.  Our nonperturbative estimates compare well with pertubative estimates of the binding energy under various coordinate transformations.  \revtext{Comparison to experimental results requires addressing the effect of virtual photon annihilation in the channel with zero orbital angular momentum.}  Along the way, we have encountered what we believe are several significant misprints in the literature upon which this work is based.  At the very least we have found some inconsistencies, as only by either removing or including some factors of 2 have we been able to get both theory and evaluation to reach a state of agreement.  We have made our programs available as a toolbox for the GNU Octave environment.  That toolbox is located at \url{https://gitlab.com/opra-ppst/opra-ppst.git}.  We include supporting derivations in the wxMaxima format.  We welcome interested readers to explore for themselves the evaluation presented here.


\backmatter

\bmhead{Supplementary information}
\revtext{
This article is accompanied by Supplemental Material in the form of files in the wxmx format to be read by wxMaxima.  These files contain details of the derivations presented here: breitJ0neg.wxmx breitJ0pos.wxmx, qed2m1\_1S0.wxmx, qed2m1\_3P0.wxmx, qed2m1\_JLS0.wxmx, qed2m1\_JLS1.wxmx, qed2m1\_PMS1.wxmx, and qed2m1.wxmx.
}

\bmhead{Acknowledgements}

The author would like to thank the members of the OPRA Association for helpful discussions and financial support to complete this project.

\begin{appendices}

\section{\label{sec:findiffcoef} Finite Difference Coefficients}
Here we present the details of our edge correction for finite difference coefficients.  Credit for publishing this algorithm belongs to~\cite{taylor2016finitedifference}, although it is probably well known in certain communities.  The goal is to calculate an approximation to the derivative operator $d^d/dx^d$ of order $d$ at any location along a finite grid $x_k \equiv k \Delta_x$ using a selection of points known as a stencil $\mbf{s}_{n,m}$ of length $n$, which is usually but not required to be odd.  The stencil is written as a row vector containing a list of integer offsets to the location at which the derivative is being evaluated, and the location of 0 offset is indicated by $m$.  In the main text we use a stencil of length $n = 7$, i.e. \beq
\mbf{s}_{7,4} = [-3, -2, -1, 0, 1, 2, 3] \;,
\eeq for points along the interior of the grid.  The order of the derivative must be less than the stencil length, $d < n$.  Next we form a column vector of integer powers from 0 to $n-1$, e.g. \beq
\mbf{p}_7 = [0, 1, 2, 3, 4, 5, 6]^T \;,
\eeq then we form the matrix of stencil values $\mbf{s}_{n,m}$ raised to powers $\mbf{p}_n$, \beq
\msf{A}_{n,m} = \mbf{s}_{n,m}^{\mbf{p}_n} \;,
\eeq which will be the LHS of a linear system of equations $\msf{A} \mbf{c} = \mbf{b}$.  On the RHS is column vector $\mbf{b}_{n,d}$ of length $n$ defined by \beq
b_k = \begin{cases}
	d\,! \;, & \mrm{if} \; k = d + 1 \;, \\
	0 \;, & \mrm{otherwise} \;,
\end{cases}
\eeq when index $k$ starts at 1.  We then solve the linear system twice (i.e. with one iteration of improvement), which we can write as a single equation, \beq
\mbf{c} = (\msf{A} \backslash \mbf{b}) - \{ \msf{A} \backslash [ \msf{A} (\msf{A} \backslash \mbf{b}) - \mbf{b} ] \} \;.
\eeq  At this point we have our set of double precision finite difference coefficients.  If the stencil is symmetric about its midpoint, as it will be for odd $n$ along the interior of the grid, we can reduce round-off error by taking the average of corresponding coefficients, \beq
\mbf{c} \leftarrow \left[ \mbf{c} + (-1)^d \mbf{c}^F \right] / 2 \;,
\eeq where $\mbf{c}^F$ has its values flipped along its axis relative to $\mbf{c}$.  An optional final step is to get the rational approximation to the coefficients $\mbf{c} = \mbf{N} / D$ where $D$ is the least common multiple of the individual denominators.  Our approximation to the derivative then includes the step size $d^d/dx^d \approx \mbf{c}_{n, m, d} / \Delta_x^d$.  For example, the approximation to $d/dx$ using a stencil $n = 5$ is \beq
\dfrac{d}{dx} \approx (12 \Delta_x)^{-1} \left[ \begin{array}{c|cccccc} -25 & 48 & -36 & 16 & -3 & 0 & 0 \\ \hline -3 & -10 & 18 & -6 & 1 & 0 & \multirow{3}{*}{\vdots} \\ 1 & -8 & 0 & 8 & -1 & 0 & \\ 0 & 1 & -8 & 0 & 8 & -1 & \\ 0 & & & \dots & & & \end{array} \right] \;,
\eeq when the boundary value $x_0$ is explicit.  In our case the function value is required to be zero at $k = 0$ and that sample is not stored, thus the first row and column do not exist.  The boundary of Table~\ref{tab:findif} is indicated by the internal lines in the matrix above.

\section{\label{sec:phisofactor2} \revtext{Examination Of Expression For $\Phi_{SO}$}}
At issue is whether the expressions for the quasipotential contributions given by Eqn.~(A28) of Appendix~A in~\cite{crater2012PhysRevD.85.116005} are truly free of typos and misprints, as claimed by Footnote 14 therein.  In particular, they give an expression for $2 \Phi_{SO}$ equal to the expression for $\Phi_{SO}$ used here.  There appears to be no argument on where Eqn.~(\ref{eqn:phiJLS1}) is supposed to land in terms of $\alpha$ so that agreement with perturbative results is maintained; the question is in how exactly one is supposed to achieve that result using the quasipotential formulation.  Note that we have observed other places in their expressions for the quasipotential terms where factors of -1 and 3 have come and gone over the years.

Let us evaluate the LHS of Eqn.(\ref{eqn:phiJLS1}) under the assumption $\tilde{\Phi}_{SO} \equiv \Phi_{SO} / 2$.  In that case, one finds \bes \begin{align}
\Phi_D - 2 \tilde{\Phi}_{SO}  & + \Phi_{SS} + 2 \Phi_T - 2 \Phi_{SOT} \\
& = \del^2 \mcal{G} + \mcal{G}'^2 - 2 \mcal{G}' / r +
[ 3 \cosh(\mcal{G}) + \sinh(\mcal{G}) ] \mcal{G}' / 2 r + [ 1 - \cosh(\mcal{G}) ] / r^2 \\
& = \frac{3 {{\alpha }^{2}}+2 r w \alpha +{{r}^{2}} {{w}^{2}}}{{{r}^{2}} {{\left( 2 \alpha +r w\right) }^{2}}}-\frac{2 {{\alpha }^{2}}+3 r w \alpha +2 {{r}^{2}} {{w}^{2}}}{2 {{r}^{\frac{5}{2}}} \sqrt{w} {{\left( 2 \alpha +r w\right) }^{3/2}}} \;,
\end{align} \ees which does not reduce to the form of Eqn.~(\ref{eqn:JLS1}) in the main text.  Similar remarks apply to Eqns.~(\ref{eqn:phi3P0}) and (\ref{eqn:3P0}) as well, where a term with $\Phi_{SO}$ also appears.

\section{\label{sec:largealpha} Large Alpha Results}
The original motivation for this work was to investigate unpublished results by Eliahu Comay concerning meson spectroscopy.  Details have been lost to time, but a manuscript from the 1970's describes using the $x$ coordinate transformation at a resolution of 100 grid points and a fifth order interpolating polynomial (stencil length of 2).  A solution for state $\prescript{1}{}S_0$ at large values of $\alpha$ with binding energy more than half of the rest mass was described.  Setting the parameters of the calculation to match that description, and neglecting the iterative improvement of the initial estimate, we can achieve similar results here.

\begin{table}
	\centering
	\begin{tabular}{|c|c|}
		\hline
		$\alpha$ & binding percentage \\
		\hline
		0.50	& 7.09 \\
		0.52	& 8.84 \\
		0.54	& 11.31 \\
		0.56	& 14.92 \\
		0.58	& 20.19 \\
		0.60	& 27.57 \\
		0.62	& 36.94 \\
		0.64	& 47.28 \\
		0.66	& 57.19 \\
		0.68	& 65.65 \\
		0.70	& 72.39 \\
		0.72	& 77.57 \\
		0.74	& 81.51 \\
		\hline
	\end{tabular}
	\caption{Binding energy for state $\prescript{1}{}S_0$ in units of percent as a function of $\alpha$}
	\label{tab:comay} 
\end{table}

In Table~\ref{tab:comay} we show results for the binding energy in units of percent $100 \times -(w - M) / M$ relative to the constituent mass of the system.  The eigenvectors have been visually inspected to ensure a consistent tracking of the ground state.  For comparison, the analytic solution at $\alpha = 0.5$ has percentage value of 7.61.  Beyond value $\alpha$ of 0.74 only the usual state at lower binding energy has been found. 

\end{appendices}


\end{document}